\documentclass[aps,reprint,twocolumn,showpacs,preprintnumbers,amsmath,amssymb,nofootinbib,superscriptaddress,showkeys]{revtex4-1}

\usepackage{slashed}
\usepackage{soul}
\usepackage{mathrsfs,bm}
\usepackage{amssymb,amsmath}
\usepackage{indentfirst}
\usepackage{graphicx,booktabs}
\usepackage{multirow}
\usepackage{overpic}
\usepackage{color}

\usepackage{hyperref}
\usepackage{graphics}
\usepackage{epsfig}
\usepackage{soul}
\usepackage{color}

\usepackage[utf8]{inputenc}

\begin{document}
\title{Auxiliary counterterms and their role in effective field theory}

\author{Manuel Pavon Valderrama}\email{mpavon@buaa.edu.cn}
\affiliation{School of Physics, Beihang University, \\ Beijing 100191, China, \email{mpavon@buaa.edu.cn}}
        
\date{\today}
\begin{abstract}
  Effective field theories include contact-range interactions (or counterterms)
  for two reasons: representing the unknown short-range physics in a model
  independent manner and ensuring the cutoff independence of observables.
  Both are intertwined: cutoff independence alone (modulo truncation errors) 
  already generates counterterms encoding physical information
  not present in the known long-range physics.
  Yet, there is also residual cutoff dependence, which is smaller
  than the uncertainties that are achievable within
  the effective field theory description and thus
  can be safely neglected in most settings.
  If one insists on exact cutoff independence though, new counterterms
  will be required, but they encode no new physical information and
  are thus what one could call redundant, or {\it auxiliary}, counterterms.
  It happens that auxiliary counterterms are still useful for solving
  certain inconsistencies that appear during renormalization or
  for improving the convergence of the effective
  field theory expansion.
  Examples of these use cases are discussed, including the interpretation of
  the improved actions or the relation between perturbative and
  non-perturbative renormalization.
\end{abstract}

\maketitle

\section{Introduction}

Effective field theories (EFTs) are generic and systematic descriptions of
low energy phenomena for which a more fundamental, high energy theory
is either unknown or impractical to solve.
Their formulation relies on the existence of a separation of scales (i.e.
a clear distinction of what constitutes low and high energy physics)
and low energy symmetries constraining the interactions
that can be built with the low energy degrees of freedom.
The quintessential example might be nuclear EFT~\cite{Bedaque:2002mn,Epelbaum:2008ga,Machleidt:2011zz,Hammer:2019poc,vanKolck:2020llt},
whose underlying theory ---
quantum chromodynamics (QCD) --- is not analytically solvable
at large distances owing to asymptotic freedom, yet there are
distinct low energy degrees of freedom (nucleons and pions)
that are not the quarks and gluons of QCD, and a symmetry,
chiral symmetry~\cite{Bernard:1995dp}, that shapes
the dynamics of these degrees of freedom.

EFTs are formulated as expansions in a ratio of scales, where observables
can be written as
\begin{eqnarray}
  \langle \hat{\mathcal{O}}_{\rm EFT} \rangle &=& \sum_{\nu = \nu_{\rm min}}^{\infty}
          {\langle \hat{\mathcal{O}} \rangle}^{(\nu)} =
          M^d\,\sum_{\nu = \nu_{\rm min}}^{\infty} c_{\nu}
          {\left( \frac{Q}{M} \right)}^{\nu} \, ,
\end{eqnarray}
with $Q$ and $M$ being the characteristic low and high energy scales of
the system one is interested in, $\langle \hat{\mathcal{O}} \rangle$
the matrix element of an operator representing a given observable, 
$\nu$ the index indicating the order in the expansion,
$d$ the energy dimensions of the matrix element and
$c_{\nu}$ dimensionless coefficients of
$\mathcal{O}(1)$ size, ideally~\footnote{For simplicity,
  the dependence of $\hat{\mathcal{O}}_{\rm EFT}$ on light scales,
  such as external momenta, has been obviated.
  This is not difficult to include, the only change being that
  the coefficients in the expansion will explicitly depend
  on light scales and their ratios, i.e. $c_{\nu} = c_{\nu} (Q, Q'/Q)$.
  Contrary to what happens with $Q/M$, this second dependence
  is in general not polynomial.
}.
The higher the order $\nu$, the more laborious the computations will be.
Yet, owing to the power series structure of EFT observables, it is convenient
to simply truncate calculations up to the desired accuracy
\begin{eqnarray}
  \langle \hat{\mathcal{O}}_{\rm EFT} \rangle &=&
  \sum_{\nu = \nu_{\rm min}}^{\nu_{\rm max}}\,{\langle \hat{\mathcal{O}} \rangle}^{(\nu)} +
  M^d \, \mathcal{O} \left( {\left( \frac{Q}{M} \right)}^{\nu_{\rm max}+1} \right)
  \, . \nonumber \\
\end{eqnarray}
Depending on how good is the separation of scales --- the $Q/M$ ratio --- and
the specific accuracy goals, it is possible to decide how many terms to keep
in the EFT expansion in advance. Or at least in principle: EFT practice often
happens to be considerably more involved than expected.

The derivation of EFTs might be better understood by the introduction of
a cutoff, a resolution scale controlling the desired level of
detail in one's own description of a physical system.
By reducing the resolution while guaranteeing that observables are independent
of the cutoff, one eventually arrives at a low energy description
that is devoid of the specific details of the (maybe unknown)
underlying theory of the system one is interested
in~\cite{Wilson:1973jj,Polchinski:1983gv}.

This idea is called renormalization group invariance (RGI) and
can be mathematically formulated as
\begin{eqnarray}
  \frac{d}{d \Lambda} \langle \Psi_{\rm EFT} | \hat{\mathcal{O}}_{\rm EFT} | \Psi_{\rm EFT} \rangle = 0 \, ,
  \label{eq:RGI}
\end{eqnarray}
where $\Lambda$ is the (momentum space) cutoff, $| \Psi_{\rm EFT} \rangle$
the EFT wave function of the system and $\hat{\mathcal{O}}_{\rm EFT}$
the EFT operator representing a given observable, where
specific applications of this idea to nuclear EFT might be found
in~\cite{Birse:1998dk,Barford:2002je,Birse:2005um,PavonValderrama:2014zeq}.
For solving the renormalization group equation (RGE) above, it is necessary
to include a series of contact-range operators in $\hat{\mathcal{O}}_{\rm EFT}$
that represent the contributions originating from the underlying theory
but that have been {\it ``integrated out''}
by the renormalization process.
Then, by evolving the RGE from $\Lambda = M$ (the unknown high energy
description) to $\Lambda = Q$ (the known low energy physics),
one determines the relative size of the contact-range
operators within the EFT.
Consequently, once this relative size is expressed in powers of the $Q/M$
ratio, then the power counting of the contributions to the EFT expansion of
the operator $\hat{\mathcal{O}}_{\rm EFT}$ is also determined.

The number of contact-range operators required to fulfill exact RGI is infinite
in principle.
While this is trivial regarding the full EFT expansion (because it has infinite
terms), what is more interesting is that this observation
also applies order-by-order.
In fact at each order in the EFT expansion it is possible to distinguish
two types of contacts~\cite{Valderrama:2016koj}:
\begin{itemize}
\item[(i)] The contacts that carry physical information (which can be
  determined from fitting observables).
\item[(ii)] The contacts that are only there to guarantee exact cutoff
  independence (and which carry no new information besides what is already
  included in the previous type of contact).
\end{itemize}
The first type are what one usually understands when referring to counterterms
in the context of renormalization as applied to the two-nucleon system
(i.e. running contact-range operators that parametrize one's own
ignorance about the details of the underlying
theory~\footnote{This is different, though related,
  to the usual quantum-field theory understanding of
  the counterterms as local terms that are included
  in the Lagrangian to make the amplitudes finite. While
  both conceptions are equivalent in the perturbative regime,
  their comparison in the non-perturbative case is open to interpretation
  (which might partly explain the different philosophies regarding
  renormalization in nuclear EFT).
}),
while the second are physically redundant and, though a component of
the internal nuts and bolts of the renormalization group flow,
bear no practical consequences in most situations.
One might call this second type redundant, or auxiliary, counterterms,
where I will favor the second term to avoid confusion
with the unrelated concept of redundant operators.

Auxiliary counterterms do in general play no practical role in most EFT settings,
except if one wants to reduce the residual cutoff dependence of calculations.
It is in this context that they appear
in pionless EFT~\cite{vanKolck:1998bw,Chen:1999tn} when regularized
with power divergence subtraction (PDS)~\cite{Kaplan:1998tg,Kaplan:1998we},
where higher order counterterms (carrying no information about new parameters
in the effective range expansion) are often included to remove
the renormalization scale dependence generated by the iteration of
lower order counterterms (which are the ones containing physical
information).
Yet, this is not exclusive to PDS: auxiliary counterterms also appear in cutoff
regularization, as discussed in~\cite{Valderrama:2016koj}, where they
were originally referred to as the redundant piece of
the contact-range couplings.
A more recent use case of auxiliary counterterms are the {\it improved actions},
i.e. the inclusion of a subset of the subleading contacts at leading order
(${\rm LO}$) as a device to improve the convergence of few-body calculations
in nuclear EFT~\cite{Contessi:2023yoz,Contessi:2024vae,Contessi:2025xue}.
In Section \ref{sec:cutoff} I will review these examples
in more detail and comment on their impact
on how one interprets cutoff dependence.

Auxiliary counterterms are also potentially useful as an analysis tool:
they can clarify and solve subtle issues that have been
previously pointed out in the literature.
In particular, Epelbaum et al.~\cite{Epelbaum:2018zli} discovered a very
intriguing inconsistency when comparing the $\hbar$ expansion of
the EFT amplitudes with the limit of said amplitudes
in the hard cutoff limit.
Besides providing a solution to the previous inconsistency, auxiliary counterterms 
also uncover a relation between the organization of the $\hbar$ expansion
and power counting.
The explanation of the inconsistency found in~\cite{Epelbaum:2018zli},
together with its solution in terms of the auxiliary counterterms
(originally explained in a comment~\cite{Valderrama:2019yiv}
to~\cite{Epelbaum:2018zli}, though in an extremely succinct manner), and
its reinterpretation as a power counting issue
in disguise will be presented in Section \ref{sec:hbar}.

The connection between perturbative and non-perturbative renormalization
is a really intriguing issue (partly because of a more
limited theoretical understanding of the latter).
In this regard,
Epelbaum et al.~\cite{Epelbaum:2018zli} have brought to the attention of
the EFT community that the non-perturbative renormalization of
singular potentials as formulated for instance
in~\cite{Beane:2000wh,PavonValderrama:2005gu,Nogga:2005hy}
might be incompatible with perturbative
renormalization (in the BPHZ
sense~\cite{Bogoliubov:1957gp,Hepp:1966eg,Zimmermann:1969jj}).
Here I argue that this incompatibility (if it exists) only appears after
taking the $r_c \to 0$ limit, while for finite cutoffs (even if they are
arbitrarily hard) there is equivalence between the two approaches.
This will be explained in detail in Section \ref{sec:pert}.

\section{Exact versus EFT Cutoff Independence}
\label{sec:cutoff}

In this section I want to revisit cutoff independence in EFT,
in particular its meaning and scope.
The cutoff is not a physical scale, rather an auxiliary one
that appears in our theories about the physical world.
However it is not necessarily devoid of physical meaning.
In Wilsonian renormalization the cutoff plays the role of
the resolution scale with which one describes the physical world.
Fundamental theories correspond to high resolutions, while EFTs are
the type of theories we obtain for low resolutions.
Yet cutoff independence is still essential to guarantee that we are describing
the same physics independently of the resolution scale.

Once one has constructed the EFT via Wilsonian renormalization
or equivalent means, the cutoff enters again into the stage.
While it is true that the full EFT computed to all orders is cutoff
independent, in actual calculations one expands the EFT according
to power counting and makes a truncation.
The expansion can break exact cutoff independence and this means that
the choice of a regularization scheme is a relevant issue.
Exact cutoff independence is not strictly necessary in this context:
if one is making a low energy expansion, physical quantities can only
be determined up to the truncation error of the expansion.
Even though in the full EFT it might be more natural to consider
the cutoff as a soft scale (after all in the Wilsonian
interpretation one is actually evolving $\Lambda \to Q$),
in actual calculations harder cutoffs might
be more convenient or even necessary.
Here I want to clarify with a few examples the issues arising
from the interplay between cutoff independence and
the EFT uncertainty.

Yet, even if one expands the EFT series only up to a finite order,
exact cutoff independence can still be implemented.
This is the case in pionless EFTs where the calculations are analytical
for some regulators.
However keeping exact cutoff independence order-by-order involves
the inclusion of the aforementioned auxiliary counterterms, which
do not contain any information about observable quantities.
These counterterms are only present to absorb the residual cutoff dependence
and, for a given dimensionality, they enter at a lower order
than counterterms encoding information about
the low energy physics.

\subsection{Pionless EFT with PDS}

A prominent example is the $C_4$ operator in pionless EFT when one regularizes
with PDS: the $C_4$ operator contains a ${\rm N^2LO}$ and ${\rm N^3LO}$
component.
The ${\rm N^2LO}$ piece cancels the residual cutoff dependence from
the iteration of the $C_2$ operator, while the ${\rm N^3LO}$ piece
is the one containing information about the shape parameter.
The argument for including a piece of $C_4$ that does not contribute
to observables is that it removes a contribution to the shape
parameter that comes from the iteration of
the $C_2$ operator.
This contribution is cutoff (or scale, if one does not want to use the term
cutoff in the context of PDS) dependent and becomes negligible
if the cutoff is hard enough.
Thus it fits the definition of a redundant or auxiliary counterterm.

It might be useful to review in more detail how this counterterm appears.
One begins by considering the contact-range potential
\begin{eqnarray}
  \langle p' | V_C | p \rangle =
  \sum_{n=0}^{\infty} \frac{C_{2n}}{2}\,(p^{2n} + {p'}^{2n}) \, ,
\end{eqnarray}
which describes the two-body potential in the absence of finite-range forces,
where $p$ and $p'$ are the incoming and outgoing momenta.
This potential generates the following on-shell T-matrix
in PDS~\cite{Kaplan:1998tg,Kaplan:1998we}
\begin{eqnarray}
  T(k) &=& -\frac{4\pi}{M_N}\,\frac{1}{k\,\cot{\delta} - i k}
  \nonumber \\
  &=&
  \frac{\sum C_{2n}(\Lambda) k^{2n}}
  {1 + \frac{M_N}{4 \pi} (\Lambda + i k)\,\sum C_{2n}(\Lambda) k^{2n}}
  \, , \label{eq:T-matrix-PDS}
\end{eqnarray}
with $M_N$ the nucleon mass (i.e. I assume the two-body system to be
the two-nucleon system) and $k$ the on-shell momentum,
where $\Lambda$ is a subtraction scale, which appears
when regularizing the loop integrals
\begin{eqnarray}
  I_0^{\rm PDS}(k; \Lambda) &=&
  \int_{\rm PDS} \frac{d^3 \vec{q}}{(2\pi)^3}\,
  \frac{M_N}{k^2 - q^2 + i\,\epsilon} \nonumber \\
  &=& - \frac{M_N}{4\pi}\,(\Lambda + i k) \, . 
\end{eqnarray}
Its function within EFT is equivalent to that of a cutoff, independently of
whether the PDS regularization does not cut the loops: at the end of
the day the PDS renormalization scale ends up being a resolution
scale, behaving exactly in the same way as one would
expect a cutoff to behave.

Instead of the T-matrix, it will be more convenient to consider
the cotangent of the phase shift, given by:
\begin{eqnarray}
  k\,\cot{\delta} =
  -\frac{4 \pi}{M_N}\,\frac{1}{\sum C_{2n}(\Lambda) k^{2n}}
  - \Lambda \, . \label{eq:kcotd-PDS}
\end{eqnarray}
The reason is that with this expression one can calculate the $C_{2n}$ couplings
by matching them with the effective range expansion (ERE) parameters
\begin{eqnarray}
  k\,\cot{\delta} = - \frac{1}{a_0} + \frac{1}{2} r_0 \, k^2 + \sum_{n=2}^{\infty}
  v_{n}\,k^{2n} \, ,
\end{eqnarray}
with $a_0$ the scattering length, $r_0$ the effective range and $v_n$
the shape parameters, yielding
\begin{eqnarray}
  C_0(\Lambda) &=& \frac{4\pi}{M_N} \frac{a_0}{1 - \Lambda\,a_0} \, , \\
  C_2(\Lambda) &=& \frac{M_N}{4\pi}\,C_0^2(\Lambda)\,\frac{r_0}{2} \, , \\
  C_4(\Lambda) &=& \frac{M_N}{4\pi}\,C_0^2(\Lambda)\,v_2 +
    \frac{C_2^2(\Lambda)}{C_0(\Lambda)} \, ,
\end{eqnarray}
plus the corresponding expressions for the higher order couplings (which
become increasingly complex).

Within pionless EFT, the $C_0$ coupling enters at ${\rm LO}$, while
the effects from the effective range $r_0$ and shape parameter $v_2$ at
${\rm NLO}$ and ${\rm N^3LO}$, respectively.
Indeed, the ${\rm LO}$ and ${\rm NLO}$ calculations reproduce
the effective range expansion exactly
\begin{eqnarray}
  k\,\cot{\delta} \Big|_{\rm LO} &=& -\frac{1}{a_0} \, , \\
  k\,\cot{\delta} \Big|_{\rm NLO} &=& -\frac{1}{a_0} + \frac{1}{2}\,r_0\,k^2 \, ,
\end{eqnarray}
and are cutoff independent.
But if one extends the calculation to ${\rm N^2LO}$ by including only
the $C_0$ and $C_2$ couplings, one obtains:
\begin{eqnarray}
  k\,\cot{\delta} \Big|_{\rm N^2LO} = -\frac{1}{a_0} + \frac{1}{2}\,r_0\,k^2 +
  \frac{{\left( \frac{1}{2}\,r_0\,k^2 \right)}^2}{\Lambda - \frac{1}{a_0}}\, ,
\end{eqnarray}
where it is apparent that there is a spurious contribution to the shape
parameter given by
\begin{eqnarray}
v_2^{\rm N^2LO} = \frac{1}{\Lambda - \frac{1}{a_0}}\,
{\left( \frac{r_0}{2} \right)}^2 \, ,
\end{eqnarray}
and whose origin is the cutoff dependence generated
by the iteration of the $C_2$ coupling.
Actually, the full ${\rm N^2LO}$ prediction for the shape parameter is basically
residual cutoff (or regularization/renormalization scale~\footnote{I will
  refer to the PDS auxiliary scale as a regularization scale, instead of its
  customary {\it ``renormalization scale''} denomination. The reason
  is that this scale by itself only regularizes the loop integrals.})
dependence.
In fact it scales as $1/\Lambda$, which means that it is indeed
a higher order contribution once one takes $\Lambda \geq m_{\pi}$
(with $m_{\pi}$ the pion mass).

However, usually in PDS the auxiliary scale $\Lambda$ is not considered
a cutoff, but a regularization scale for which no judgment
about its size is held.
While from the point of view of cutoff regularization one might leave things
as they are and consider the ${\rm N^2LO}$ shape parameter as part of
the EFT uncertainty, within PDS the instinct is to remove this ambiguity.
For this, one calculates the $C_4$ coupling, where it is apparent that
there are two contributions of a very different nature:
\begin{eqnarray}
  C_4(\Lambda) &=&
  \underbrace{\frac{M_N}{4\pi}\,C_0^2(\Lambda)\,v_2}_{{\rm N^3LO}} +
  \underbrace{\frac{C_2^2(\Lambda)}{C_0(\Lambda)}}_{{\rm N^2LO}} \, .
\end{eqnarray}
The part of $C_4$ encoding new physical information --- the shape parameter ---
enters at the order at which one expects $v_2$ to enter: ${\rm N^3LO}$.
Meanwhile, the part of $C_4$ that does not encode new physical information
(the auxiliary component of $C_4$), but instead represents the iteration of
the $C_2$ coupling, enters at the order at
which this iteration does: ${\rm N^2LO}$.
That is, at the order at which the residual cutoff dependence appeared.

It is interesting to notice that one might arrive to the same conclusion
by considering the following formalism to define the power counting.
For this, I begin with the set of light scales $Q$, which for pionless EFT
is given by
\begin{eqnarray}
  Q = \{ k, \frac{1}{a_0} \} \, ,
\end{eqnarray}
where $k$ is the external momentum and $a_0$ the scattering length (while
for other EFTs this set will be different~\footnote{For instance, 
  if all two-body contact-range interactions are perturbative,
  then $Q = \{ k \}$. If one considers a pionless EFT in
  which the effective range $r_0$ has to be resummed,
  then $Q = \{ k, \frac{1}{a_0}, \frac{1}{r_0} \}$.
  If $a_0$ and $v_2$ scales as $1/Q$, but $r_0$ is natural,
  then $Q = \{ k, \frac{1}{a_0}, \frac{1}{M^2\,v_2} \}$,
  where the factors of $M$ have been included to write
  the set Q in the correct dimensions, i.e. $v_2 \sim 1/(M^2 Q)$.
  But if one simply worries about where a factor of $Q$ is contained,
  dimensional correctness might be waived (after all, this is
  merely a notation device), in which case
  one may simple write $Q = \{ k, \frac{1}{a_0}, \frac{1}{v_2} \}$.
}).
To find the power counting expansion of a given physical observable, operator
or any other quantity of interest one simply rescales $Q \to \lambda Q$:
\begin{eqnarray}
  \mathcal{O}(\lambda Q) = \sum_{\nu}\,\lambda^{\nu} \,\mathcal{O}^{(\nu)}(Q) \, ,
\end{eqnarray}
with $\mathcal{O}$ the physical object we are interested in, and where
a contribution of order $Q^{\nu}$ is now multiplied by
$\lambda^{\nu}$.
That is, by means of rescaling counting becomes automatic.
A trivial example within pionless EFT is the effective range expansion,
for which rescaling yields 
\begin{eqnarray}
  k\,\cot{\delta} = - \frac{\lambda}{a_0} + \frac{\lambda^2}{2} r_0 \, k^2 +
  \sum_{n=2}^{\infty}\,\lambda^{2n}\,v_{n}\,k^{2n} \, ,
\end{eqnarray}
where it is apparent that the scattering length $a_0$ enters at ${\rm LO}$
(the lowest scaling), the effective range $r_0$ at ${\rm NLO}$ and
the n-th shape parameter $v_n$ at ${\rm N^{2n-1}LO}$.

The scaling properties of the cutoff, which are necessary for calculating
the power counting of the couplings, can be derived
from the unregularized loop integral
\begin{eqnarray}
  I_0(k) &=& \int \frac{d^3\,\vec{q}}{(2\pi)^3}\,G_0(E) \nonumber \\
  &=&
  \int \frac{d^3\,\vec{q}}{(2\pi)^3}\,\frac{M_N}{k^2 - q^2 + i \epsilon} \, ,
\end{eqnarray}
where by rescaling $Q \to \lambda Q$ one finds
\begin{eqnarray}
  I_0(k) \to I_0(\lambda k) = \lambda I_0(k) \, , 
\end{eqnarray}
which is basically the well-known result that each iteration
induces a factor of $Q$ in the power counting.
If one introduces a cutoff of any type and requires that the presence of
a cutoff does not modify the scaling properties of the unregularized
loop integral, the conclusion is that the cutoff
is effectively a light scale:
\begin{eqnarray}
  I_0(k; \Lambda) \to
  I_0(\lambda k; \lambda \Lambda) = \lambda I_0(k; \Lambda) \, .
\end{eqnarray}
The interesting thing is that this applies either to a cutoff actually
cutting the ultraviolet divergence of the loop integral in cutoff
regularization or to the dimensional rescaling of the couplings
that one introduces in PDS before removing the poles
at $D=3$ and $4$ space-time dimensions.

With the addition of the rescaling rule $\Lambda \to \lambda\,\Lambda$,
the first two couplings behave as
\begin{eqnarray}
  C_0(\lambda \Lambda; \frac{a_0}{\lambda}) &=&
  \frac{1}{\lambda}\,\frac{4\pi}{M_N} \frac{a_0}{1 - \Lambda\,a_0} \, , \\
  C_2(\lambda \Lambda; \frac{a_0}{\lambda}) &=& \frac{1}{\lambda^2}\,
  \frac{M_N}{4\pi}\,C_0^2(\Lambda; a_0)\,\frac{r_0}{2} \, , 
\end{eqnarray}
which reproduce the counting of $C_0$ and $C_2$, i.e. that they are
promoted by a factor of $Q^{-1}$ and $Q^{-2}$ with respect to
their size in naive dimensional analysis.
Notice that the dependence in the scattering length has been made explicit
in the expressions above to make the analysis more transparent.
By applying the same rescaling rules to $C_4$ one finds
\begin{eqnarray}
  C_4(\lambda \Lambda, \frac{a_0}{\lambda}) &=&
  \frac{1}{\lambda^2}\,\frac{M_N}{4\pi}\,C_0^2(\Lambda)\,v_2 +
  \frac{1}{\lambda^3}\,\frac{C_2^2(\Lambda)}{C_0(\Lambda)} \, ,
\end{eqnarray}
which indicates that the redundant part of $C_4$ receives a $Q^{-3}$ promotion,
while for the part of $C_4$ containing physical information
the promotion is the expected $Q^{-2}$.

The bottom-line though is that a cutoff or a regularization scale are
functionally indistinguishable at the EFT level: the decision to leave
the residual cutoff dependence as it is or to remove it completely
corresponds more to a philosophy or attitude than to anything
essential about the regularization process.
One might ignore the auxiliary counterterms in PDS, or one might decide to
take them seriously in cutoff regularization,
which is what I will do next.

\subsection{Pionless EFT with a cutoff}

In the case of a pionless theory with cutoff regularization,
one can argue along similar lines as in PDS.
For a contact-range potential with a delta-shell regulator
\begin{eqnarray}
  V_C(r; r_c) = C_k(r_c)\,\frac{\delta(r-r_c)}{4 \pi r_c^2} \, ,
\end{eqnarray}
the exact solution of the pionless EFT in the two-nucleon sector
is given by solving the equation
\begin{eqnarray}
  k\,\cot{(k r_c + \delta)} - k\,\cot{k r_c} =
  \frac{M_N}{4\pi r_c^2}\,C_k(r_c) \, ,
  \label{eq:Ck-boundary}
\end{eqnarray}
from which one arrives at
\begin{eqnarray}
k\,\cot{\delta} &=& \frac{k + \left( \frac{M_N}{4\pi r_c^2}\,C_k(r_c)
+ k\cot{k r_c}\right)\,\cot{k r_c}}
{-\frac{1}{k}\frac{M_N}{4\pi r_c^2}\,C_k(r_c)}
\, . \nonumber \\ \label{eq:kcotd-delta-shell}
\end{eqnarray}
In the equations above $C_k(r_c)$ stands for the expansion
\begin{eqnarray}
C_k(r_c) = \sum_{n = 0}^{\infty} C_{2n}(r_c; a_0)\,k^{2n} \, ,
\end{eqnarray}
where I have explicitly indicated the dependence of the $C_{2n}$'s
on the scattering length (and, of course, the cutoff radius).
The rationale is that this choice makes it explicit the power counting
expansion of the $C_{2n}$ couplings.
If the low-energy scales are
\begin{eqnarray}
Q = \{ \frac{1}{a_0}, k \} \, ,
\end{eqnarray}
by rescaling ($Q \to \lambda\,Q$ and $r_c \to r_c / \lambda$) one finds that
\begin{eqnarray}
C_{2n}(\frac{r_c}{\lambda}; \frac{a_0}{\lambda}) =
\sum_{\nu} \,\lambda^{\nu}\,C_{2n}^{(\nu)}(r_c; a_0) \, ,
\end{eqnarray}
where the power counting is explicit ($\lambda^{\nu}$ 
corresponds to $Q^{\nu}$).
One can write then the first few $C_{2n}$ couplings as
\begin{eqnarray}
C_0(\frac{r_c}{\lambda}; \frac{a_0}{\lambda}) =
\frac{4 \pi r_c^2}{M_N} &\Big[& \frac{1}{\lambda}\,
\frac{a_0}{r_c \, (r_c - a_0)} \Big] \, , \\
C_2(\frac{r_c}{\lambda}; \frac{a_0}{\lambda}) =
\frac{4 \pi r_c^2}{M_N} &\Big[&
\frac{1}{\lambda^3}\,\frac{a_0 \, r_c\,(r_c - 2\,a_0)}{3\,(r_c - a_0)^2}
\nonumber \\
&+& \frac{1}{\lambda^2}\,\frac{a_0^2\,r_0}{2 \, (r_c - a_0)^2} \Big] \, ,
\label{eq:C2-scaling-rc} \\
C_4(\frac{r_c}{\lambda}; \frac{a_0}{\lambda}) =
\frac{4 \pi r_c^2}{M_N} &\Big[&
\frac{1}{\lambda^5}\,
\frac{a_0 \, r_c^3\,(14 a_0^2  - 12 a_0\,r_c + 3 r_c^2)}{(r_c - a_0^3)}
\nonumber \\
&+& \frac{1}{\lambda^4}\,
\frac{a_0^2 \, r_c^2\,(r_c - 3 a_0)\,r_0}{6\,(r_c - a_0)^3}
\nonumber \\
&+& \frac{1}{\lambda^3}\,\frac{a_0^3 \, r_c\,r_0^2}{4\,(r_c - a_0)^3}
\nonumber \\
&+& \frac{1}{\lambda^2}\,\frac{a_0^2 \, v_2}{(r_c - a_0)^2} \Big]
\, , 
\end{eqnarray}
plus the solutions for the higher order $C_{2n}$'s.

The form in which the $C_{2n}$'s are written highlights that the $C_{2n}$'s
contain a contribution scaling as $1/\lambda^{2n+1} \sim 1/Q^{2n+1}$.
If one takes into account that the contact-range operator is actually
the combination $C_{2n}\,k^{2n}$, one finds that the lowest order
contribution is always of order $Q^{-1}$.
That is, all contact-range operators contain a piece that enters at ${\rm LO}$.
However, it is important to notice that the ${\rm LO}$ piece only contains
physical information in the particular case of the $C_0$ coupling.
For all the other couplings the ${\rm LO}$ piece is only there to counter
the cutoff dependence of the ${\rm LO}$ phase shift.
Actually the first non-trivial piece of the $C_{2n}$ coupling that contains
new physical information, i.e. that fixes a new low-energy constant,
scales as $1/Q^2$ and enters at $\rm N^{2n-2}LO$ ($Q^{2n-2}$),
as expected in pionless EFT.

At this point one has two obvious options for organizing the EFT expansion.
The first is the standard one in pionful EFT: one includes the $C_{2n} k^{2n}$
contact-range operator at the order for which it fixes
a new low-energy constant.
This amounts to sacrificing exact cutoff independence.
Yet one recovers cutoff independence in the ultraviolet limit. Thus one might
informally refer to this approach as ultraviolet cutoff independence.
In this case the ${\rm LO}$ contact-range potential is given by
\begin{eqnarray}
\frac{M_N \,C_k^{\rm LO}(r_c)}{4 \pi r_c^2} &=&
\frac{a_0}{r_c \, (r_c - a_0)} \, ,
\end{eqnarray}
while the ${\rm LO}$ amplitude is
\begin{eqnarray}
k\,\cot{\delta_{\rm LO}} &=& - k\,
\frac{2 k r_c (r_c - a_0) + a_0\,\sin{2 k r_c}}{2 a_0 \sin^2(k r_c)} \, .
\end{eqnarray}
After expanding in terms of $k^2$ one has
\begin{eqnarray}
\label{eq:ERE-LO}
k\,\cot{\delta_{\rm LO}} &=& - \frac{1}{a_0} +
\left( \frac{2 r_c}{3} - \frac{r_c^2}{3 a_0}\right) \, k^2 + \dots
\end{eqnarray}
from which one sees that the absolute cutoff dependence is
$\mathcal{O}(k^2\,r_c)$ and the relative one (i.e. after removing the $k$ factor
that is also present in $k\,\cot{\delta}$) is $\mathcal{O}(k\,r_c)$.
One expects the relative error of the ${\rm LO}$ phase shift
to be $\mathcal{O}(k/m_{\pi})$, which means that one has to choose a cutoff
of the order of $m_{\pi}$ (or harder) to ensure that the cutoff dependence
is at most of the size of the EFT error.
For the cutoff radius this translates into the condition $r_c \leq 1/m_{\pi}$.

The second option is to have exact cutoff independence.
In principle this looks a hopeless task due to the infinite number of
counterterms that are required to counter the cutoff dependence.
However pionless EFT with a delta-shell regulator is analytically solvable.
If one uses the following ${\rm LO}$ contact-range coupling
\begin{eqnarray}
\frac{M_N \,C_k^{\rm LO}(r_c)}{4 \pi r_c^2} &=& - k\,\cot{k r_c}
- \frac{k\,\left( 1 + \frac{\cot{k r_c}}{a_0 k} \right)}
{\cot{k r_c} - \frac{1}{a_0 k}} \, ,
\end{eqnarray}
one ends up with the phase shift
\begin{eqnarray}
k\,\cot{\delta_{\rm LO}(r_c)} = - \frac{1}{a_0} \, ,
\end{eqnarray}
for all $r_c$ (i.e. there is exact cutoff independence).

But the previous strategy, though interesting, has a flaw in its design.
Exact cutoff independence, that is
\begin{eqnarray}
\frac{d}{d r_c} k \, \cot{\delta_{\rm LO}(r_c)} = 0 \, ,
\end{eqnarray}
does not imply the existence of a privileged solution,
namely the $r_c = 0$ one.
One can actually choose the solution for a reference cutoff radius $r_c = R$
as the ${\rm LO}$ phase shift
\begin{eqnarray}
k\,\cot{\delta_{\rm LO}} &=& - k\,
\frac{2 k R (R - a_0) + a_0\,\sin{2 k R}}{2 a_0 \sin^2(k R)} \, ,
\label{eq:kcot-alt}
\end{eqnarray}
which implies the ${\rm LO}$ contact-range potential:
\begin{eqnarray}
\frac{M_N \,C_k^{\rm LO}(r_c; R)}{4 \pi r_c^2} &=& 
\frac{\frac{a_0\,k}{\sin^2(k r_c)}}{\frac{k\,(R-a_0)\,R}
  {\sin^2(k R)} + a_0\,(\cot{k R} - \cot{k r_c})} \, . \nonumber \\
\label{eq:Ck-alt}
\end{eqnarray}
The reference cutoff $R$ is only subjected to the condition
\begin{eqnarray}
\label{eq:condition-LO}
\left| k \, \cot{\delta_{\rm F}} - k \, \cot{\delta_{\rm LO}(R)} \right| \leq 
\mathcal{O}(\frac{Q}{M}) \, ,
\end{eqnarray}
with $\delta_F$ the full (or physical) phase shift.
That is, the ${\rm LO}$ and physical phase shifts
are the same modulo the EFT uncertainty.

\subsection{The cutoff as a light scale}

The cutoff is not a physical scale, but an auxiliary one.
It is a property of a type of theories --- effective field theories ---
but not of nature.
Thus, {\it a priori}, there is no preferred interpretation regarding
its size in comparison to the physical scales.

Yet, {\it a posteriori}, it turns out though that it {\it formally}
behaves as a light scale.
The emphasis is on the formal character of this identification:
this is merely a characteristic of the EFT formalism presented
here that does not necessarily have anything to do
with the preferred or ideal size of
the cutoff in actual calculations.

This interpretation is however in conflict with the more common
one of treating the cutoff as a hard scale.
To understand the differences between them, one may revisit
the ${\rm LO}$ calculation of $\cot{\delta}$ (dimensionless,
which makes the discussion about EFT convergence
more transparent) with a finite cutoff
\begin{eqnarray}
  \cot{\delta}_{\rm LO} &=& - \frac{1}{k a_0}
  + \left( \frac{2 r_c}{3} - \frac{r_c^2}{3\,a_0}\right) \, k
  + \dots \, , 
\end{eqnarray}
where the dots represent higher powers of the momentum $k$.
If the cutoff is a light scale, the ${\rm LO}$ calculation scales
homogeneously as
\begin{eqnarray}
  \cot{\delta}_{\rm LO} (\lambda\,k , \frac{a_0}{\lambda}, \frac{r_c}{\lambda})
  = \cot{\delta}_{\rm LO} (k, {a_0}, {r_c}) \, ,
\end{eqnarray}
which implies that the cutoff dependence is also ${\rm LO}$ in principle.
That is, the calculation by itself is ambiguous (a point that will be clearer
later when I consider the interpretation of the cutoff as a hard scale).
But the cutoff is not physical, and there is the additional condition
that a ${\rm LO}$ calculation differs from the experimental or full
$\cot{\delta}_F$ by less than the expansion parameter
\begin{eqnarray}
  \left| \cot{\delta}_F - \cot{\delta_{\rm LO}}(r_c) \right| \leq
  \mathcal{O}\left( \frac{Q}{M} \right) \, ,
\end{eqnarray}
which at low momenta is equivalent to the condition
\begin{eqnarray}
  \left|  \frac{2\,r_c}{3} - \frac{r_c^2}{3\,a_0}- \frac{r_0}{2} \right|
  \leq \mathcal{O}\left( \frac{1}{m_{\pi}} \right) \, ,
\end{eqnarray}
where I have explicitly taken into account that for pionless $M = m_{\pi}$.
At the order-of-magnitude level, the previous condition is fulfilled for
\begin{eqnarray}
  r_c \leq \mathcal{O}(r_0, \frac{1}{m_{\pi}}) \, ,
\end{eqnarray}
which further reduces to $r_c \leq \mathcal{O}(1/m_{\pi})$ once one notices
that the size of the effective range is expected to be natural in units
of the Compton wavelength of the pion,
i.e. $m_{\pi} r_0 \sim \mathcal{O}(1)$. 

If the cutoff is treated as a hard scale instead, the scaling of the ${\rm LO}$
calculation would behave as
\begin{eqnarray}
  \cot{\delta}_{\rm LO} (\lambda\,k , \frac{a_0}{\lambda}, {r_c})
  = - \underbrace{\frac{1}{a_0 k}}_{Q^0} +
  \lambda\,\underbrace{\frac{2 r_c}{3}\,k}_{Q^1} + \dots \, ,
\end{eqnarray}
where the dots contain higher powers of $k$ or $\lambda$.
The underbraces indicate the order of the contribution to the calculation,
which implies that if the cutoff is not a light scale, then the cutoff
dependence of $\cot{\delta}_{\rm LO}$ is a higher order effect
beginning at ${\rm NLO}$.
Therefore, the $r_c \to 0$ limit is special as it represents
the only part of the calculation that is genuinely ${\rm LO}$.

If this is the case one can choose a suitable cutoff
from the following condition
\begin{eqnarray}
  \left| \cot{\delta_{\rm LO}(r_c \to 0)} -
  \cot{\delta_{\rm LO}}(r_c) \right| \leq
  \mathcal{O}\left( \frac{Q}{M} \right) \, ,
\end{eqnarray}
which captures the idea that the cutoff dependence is subleading and hence
the $r_c \to 0$ limit provides the pure ${\rm LO}$ piece of the calculation.
This condition is equivalent to $r_c \leq \mathcal{O}(1/m_{\pi})$, which turns
out to be exactly the same condition one obtains when considering
the cutoff as a light (but unphysical) scale.
That is, regardless of whether the cutoff is formally treated as a light
scale or not, at the end it is constrained by the condition of
convergence of the EFT series (unless the calculation
has been arranged to be exactly cutoff independent
from the start).

The bottom-line here is that the cutoff is not a physical parameter,
but a parameter of the theory.
Whether it is considered formally a light or a hard scale is immaterial 
for practical calculations, though it is important
for the inner workings of the EFT.
For instance, pionless EFT has been formulated both in PDS~\cite{Chen:1999tn}
(which treats the cutoff or regularization scale as a light scale) and
in the usual cutoff approach in which the cutoff is at least of
the order of the hard scale~\cite{vanKolck:1998bw}.
Usually EFT formulations using PDS achieve exact cutoff independence, which
leads them to choose a PDS regularization scale (or renormalization scale,
as is usually referred to in the literature) of the order of
the light scales, which has the advantage of making the analysis of
the power counting of the contact-range couplings more transparent.
Yet, as previously shown, one might as well abandon exact cutoff independence
within PDS (by not including the auxiliary counterterms) and choose
a regularization scale of the order of the hard scale,
which leads to the expected EFT convergence patterns.
Conversely one might modify the usual finite cutoff formulation as to generate
cutoff independent amplitudes, which would perfectly allow for cutoffs of
the order of the soft scale, though this is in general very difficult to
accomplish at the practical level except for very specific regularization
schemes (such as delta-shell regularization).

It is worth noticing though that while the previous example suggests that
both interpretations of the cutoff lead to the same power counting,
the discussion in~\cite{Epelbaum:2017byx} proposes two different
possibilities of organizing the power counting within
a non-relativistic EFT that exactly differ
on the point of whether the cutoff is formally a light or a hard scale.
There it is argued that counting the loops as $Q^0$, which is equivalent ---
up to a certain extent --- to counting the cutoff as a hard scale,
leads to a Weinberg-style counting in pionless EFT, instead of
the usual counting derived in~\cite{vanKolck:1998bw,Chen:1999tn}.
In contrast, counting the loops as $Q$ results in the usual counting.
This discrepancy is very intriguing and probably merits further
investigation in the future.

Independently of whether both interpretations are ultimately
equivalent~\cite{vanKolck:1998bw,Chen:1999tn} or not~\cite{Epelbaum:2017byx}
(or whether the discrepancy originates from different implicit
assumptions that are not evident at first sight),
there is an interesting advantage though
in treating the cutoff as a light scale.
This advantage lies in the coherent power counting description of observable
and non-observable EFT quantities.
The application of the rescaling rules (where $Q_{\Lambda}$ or $Q_{r_c}$ is the
set of the light scales $Q$ plus the cutoff):
\begin{eqnarray}
  Q_{\Lambda} \to \lambda\,Q_{\Lambda}
  \quad \mbox{or} \quad
  Q_{r_c} \to \lambda\,Q_{r_c} \, ,
\end{eqnarray}
automatically determines the counting of observables and the anomalous
dimension of the contact-range couplings, which in turn gives
their counting~\cite{PavonValderrama:2014zeq}.
Meanwhile, if the cutoff is formally treated as a hard scale, the analysis of
the power counting of the contact-range couplings becomes more laborious.

\subsection{The cutoff as the EFT convergence dial}

One of the most interesting and surprising implications of enforcing
order-by-order cutoff independence (when the cutoff is formally treated
as a light scale) is that while cutoff independence can indeed be achieved,
there is no intrinsic reason why a particular
value of the cutoff is better than other.
The criterion by which certain cutoffs are to be preferred is extrinsic and
given by the condition of reproducing observable quantities
within the EFT truncation errors.
This leads to a paradoxical conclusion in the sense that cutoff dependence
creeps back into the picture simply because any cutoff providing calculations
accurate up to the expected EFT uncertainty can act as a reference cutoff.
At the end genuine cutoff independence is only achievable
when considering the full EFT expansion to all orders.

This is a bit different than the usual idea of ultraviolet cutoff independence,
where there is a bias towards favoring the $r_c \to 0$ limit over the comparison
with observables.
This bias is justifiable though if the cutoff is not
formally treated as a light scale,
but as a hard scale.
However the existence of an inherent EFT uncertainty qualifies
the meaning of cutoff independence and extends
the range of acceptable cutoffs.
That is, one is entitled to choose the cutoff radius in such a way
that the ${\rm LO}$ phase shift is equivalent
to the physical one within the EFT error.
But not only that: indeed one is further entitled to minimize this error.

The first of these two statements, though equivalent at the order-of-magnitude
level to the comparison to the $r_c \to 0$ limit, allows for a larger set of
acceptable cutoffs once concrete values are considered.
To illustrate this difference one might consider the ${\rm LO}$ prediction
for the effective range in Eq.~(\ref{eq:ERE-LO}) and assume
the $r_c / a_0$ factors to be negligible.
By comparing this ${\rm LO}$ calculation with its own $r_c \to 0$ limit,
one finds the condition
\begin{eqnarray}
\frac{2\,r_c}{3} \leq \mathcal{O}\left( 
\frac{1}{m_{\pi}} \right) \, ,
\end{eqnarray}
while a direct comparison with the physical effective range gives
\begin{eqnarray}
\left| \frac{2\,r_c}{3} - \frac{r_0}{2} \right| \leq \mathcal{O}\left( 
\frac{1}{m_{\pi}} \right) \, ,
\end{eqnarray}
which is considerably less restrictive.
The reason is that $m_{\pi} r_0 = 1.87$ and $1.23$ for the singlet
and triplet, respectively, which for the second condition
has to be added to the $\mathcal{O}(1)$ numerical
factors already present in the first one, hence
resulting in a larger range of cutoffs.

It is thus worth emphasizing that the comparison to observables
is a better option for choosing a cutoff than
the naive perceptions about its ideal size:
while the first is an objective criterion, the second is subjected
to the existence of unknown numerical factors.
This latter problem is exacerbated in nuclear physics
as a consequence of the poor separation of scales.

The second statement --- that one can choose the cutoff radius
as to generate EFT observables as close to the physical ones as one wants ---
does not break any of the tenets of EFT,
yet provides better convergence.
It is important to stress that the EFT truncation error is not going to change
regardless of whether the difference between the calculated and physical
results is considerably smaller than this truncation error.
But by choosing this type of privileged cutoff, the subleading order
corrections are expected to be smaller (the truncated calculation will
require less corrections).
As a consequence, the EFT expansion will converge faster
for these cutoffs.

There are two conditions that must be met though before dialing the cutoff
for convergence: (i) there must be at least some sort of cutoff
independence in the first place, and (ii) the EFT must be convergent.
Otherwise using a favorable cutoff will be phenomenology instead of EFT.
The point is that the freedom to choose a good cutoff is only a legitimate
move once any uncontrollable hard cutoff behavior is absent.

Actually these ideas are anything but new.
The interpretation of the regularization/renormalization scale (i.e. the cutoff)
dependence not as a fatal ambiguity but rather as a parameter
that controls the convergence is standard in perturbative QCD,
where different criteria have been proposed for selecting
the scale~\cite{PhysRevD.23.2916,PhysRevD.28.228}.
Within the context of nuclear EFT, Barford and Birse already proposed
this interpretation in Ref.~\cite{Barford:2002je}. 
Later this idea played a central role in Ref.~\cite{Beane:2008bt},
which attempted to check whether the KSW counting could converge
with a more lenient choice of regulator (though unfortunately
it seems that this does not help~\cite{Gegelia:priv}).

\subsection{Improved actions and their interpretation}

An interesting aspect of the previous ideas about cutoff dependence
is their connection
with the {\it improved actions} that have been recently
proposed~\cite{Contessi:2023yoz,Contessi:2024vae,Contessi:2025xue}
as a tool to enhance the convergence of pionless EFT,
particularly in the few-body sector. 

In this context an improved action refers to the inclusion at ${\rm LO}$ of
interactions that in principle enter at subleading orders to improve
specific convergence properties of the EFT.
The difference with a modification of the power counting lies in the fact that
subleading order effects (in the original counting) are eventually able
to fully compensate for the additional ${\rm LO}$ interactions (while
a modification of the counting usually entails further
changes at higher orders).

In this regard if one considers that there is no privileged cutoff a priori
--- but that the choice of the cutoff is made a posteriori instead ---
then the cutoff can be calibrated as to approximate the full or
physical observables beyond the EFT truncation error.
If one considers the specific example of the effective range,
one chooses $r_c = R$ such that
\begin{eqnarray}
  \frac{2\,R}{3} - \frac{R^2}{3\,a_0} = x\,\frac{r_0}{2} \, ,
  \label{eq:improved-cutoff}
\end{eqnarray}
where $0 \leq x \leq 1$, depending on what fraction of the physical effective
range one wants to reproduce.
This cutoff could be referred to as an {\it improved cutoff}.

Yet, instead of tuning the cutoff, one is also entitled to exploit
that (i) there is no privileged cutoff, and that (ii) it is possible
to make use of the auxiliary counterterms to construct a contact-range
coupling reproducing part of the effective range but
that leads to cutoff independence.
For the delta-shell regulator the closed form of such a contact-range coupling
is given by Eq.~(\ref{eq:Ck-alt}), which generates
the ${\rm LO}$ amplitude of Eq.~(\ref{eq:kcot-alt}).
By setting $R$ as in Eq.~(\ref{eq:improved-cutoff})
one ends up with an improved action.

It is worth noticing though that the coupling in Eq.~(\ref{eq:Ck-alt}) does
not only fix the effective range (to a fraction of its physical value)
but the whole amplitude.
If one is interested in the inclusion of a ${\rm LO}$ coupling that only
fixes the effective range, it is enough to expand the coupling in
Eq.~(\ref{eq:Ck-alt}) in powers of $k^2$, in which case one obtains
the following $C_2$ coupling
\begin{eqnarray}
  \frac{M_N \,C_2^{\rm LO}(r_c; R)}{4 \pi r_c^2} &=&
  \frac{a_0 \, r_c\,(r_c - 2\,a_0)}{3\,(r_c - a_0)^2} \nonumber \\
  &+&
  \frac{2 R}{3 a_0}(2\,a_0 - R)\,\frac{a_0^2}{2 \, (r_c - a_0)^2} \, ,
\end{eqnarray}
which generates $r_0^{\rm LO} = x\,r_0$ for $R$ fulfilling
Eq.~(\ref{eq:improved-cutoff}).
In general, for most regulators, the fully cutoff independent
construction of Eq.~(\ref{eq:Ck-alt}) is not achievable
at the practical level.
Instead one is usually limited to include $n$ additional ${\rm LO}$ contacts
removing the cutoff dependence up to $k^{2n+2}$ terms in the ERE.
Indeed, most improved actions
in~\cite{Contessi:2023yoz,Contessi:2024vae,Contessi:2025xue} 
are limited to the inclusion of part of the effective range
(though the formalism in~\cite{Contessi:2024vae} could
be easily extended to the resummation of other ERE parameters).

The ideas of cutoff ambiguity and auxiliary counterterms imply that
improved actions are actually equivalent to tuning
the cutoff as to reproduce part of
the physical effective range.
What changes is the interpretation of why improving the ${\rm LO}$ calculation
is compatible with the EFT principles:
\begin{itemize}
\item[(i)] In~\cite{Contessi:2023yoz,Contessi:2024vae,Contessi:2025xue}
  the idea is that there is no difference between improved and non-improved
  calculations at higher orders, as the subleading operators will
  be able to reabsorb any changes made to the ${\rm LO}$ action.
\item[(ii)] But if there is no privileged cutoff (i.e. once the ${\rm LO}$
  calculation coincides with the full or experimental one
  within truncation errors), then using the cutoff ambiguity
  already present at lower orders as to improve the convergence of
  the calculation is perfectly okay, as the higher order
  calculations will display increasingly smaller cutoff
  dependence anyways.
\end{itemize}
Alternatively, in the second interpretation, one might include auxiliary
counterterms instead, which is basically a different name
for the improved actions.
If one considers that the cutoff does not (formally) belong to the light
scales, then the only valid interpretation
will be the first.
Yet, it is worth noticing that independently of this, both interpretations
are indistinguishable at the practical level.
One must not forget though that the aim of improved actions is
the enhanced convergence of the EFT series
in few-body calculations,
while the present work is limited to the two-body sector.

\subsection{Convergence around a bound state}

The expansion of the scattering amplitude around a bound state is a corner
case in which the convergence expectations derived
from power counting are not followed.
Yet, convergence might be restored by either of these two methods:
(i) the choice of a suitable cutoff or (ii) the construction of an
improved action.
Both are in turn equivalent to the well-known idea of using the bound state
pole of the scattering amplitude as the on-shell renormalization
point~\cite{Phillips:1999hh}.

If one considers a contact-range theory with a bound state (where
for the moment the $r_c \to 0$ limit will be taken), but in which
the parameters of the EFT are initially determined from the low-energy
scattering observables (i.e. the ERE parameters), then the location of
the bound state is not fixed at ${\rm LO}$ but is expressed
instead as an expansion.
In this setup the bound state equation reads
\begin{eqnarray}
  \left( k\,{\rm cot}{\delta} - i\,k \right)\Big|_{k = i \gamma_p} = 0 \, ,
\end{eqnarray}
with $k\,{\rm cot}{\delta}$ given by the ERE.
By solving this equation the expansion of $\gamma_p$ in
the $Q = \{ 1/a_0, k \}$ counting is
\begin{eqnarray}
  \gamma_p &=& \frac{1}{a_0} + \frac{r_0}{2 a_0^2}
  + \frac{r_0^2}{2 a_0^3} + \left( \frac{5}{4}\,
  \frac{r_0^3}{2\,a_0^4} - \frac{v_2}{a_0^4} \right)
  \nonumber \\
  &+& M\,\mathcal{O}\left( {\left( \frac{Q}{M} \right)}^5 \right)  \, ,
\end{eqnarray}
where $\gamma_p$ is the wave number of the pole of the scattering amplitude,
with $\gamma_p > 0$ ($\gamma_p < 0$) for a bound (virtual) state.

While the EFT expansion of $k\,\cot{\delta}$ follows the expected power
counting, this is not the case near a bound state.
The EFT expansion near the pole of the on-shell $T$-matrix is
\begin{eqnarray}
  T(k = i\,\gamma) &\to& 
    -\frac{4\pi}{M_N}\,\frac{a_S^2}{\gamma - \gamma_p} \, ,
\end{eqnarray}
for $\gamma \to \gamma_p$ , with the residue $a_S^2$ given by
\begin{eqnarray}
  \frac{1}{a_S^2} = 1 - \gamma_p\,r_0 +
  \sum_{n=2}^{\infty}\,{(-1)}^{n}\,2n\,\gamma_p^{2n-1}\,v_{n}
  \, .
\end{eqnarray}
By expanding close to the pole, one obtains
\begin{eqnarray}
  \frac{a_S^2}{\gamma - \gamma_p} = \frac{1}{\gamma - \gamma_p^{\rm LO}}
  + \frac{2\,\delta a_S}{\gamma - \gamma_p^{\rm LO}} +
  \frac{\delta \gamma_p}{{(\gamma - \gamma_p^{\rm LO})}^2} + \dots \, ,
  \nonumber \\
\end{eqnarray}
where $\delta a_S$ and $\delta \gamma_p$ refer to the subleading order
contributions to the residue and the pole's wave number.
It is apparent in this expansion that the subleading correction coming from
the change in the pole position becomes uncontrollably large
in the vicinity of $\gamma = \gamma_p^{\rm LO}$.
In particular one obtains a double pole, which is a beautiful
illustration of the instability of the infrared fixed point
corresponding to unnatural scattering lengths, whose
perturbations scale as $Q^{-2}$~\cite{Birse:1998dk}.
This convergence problem also aligns with the observations
in~\cite{Yang:2023wci}, though there the problem is
expressed in terms of the behavior of eigenvalues
in the harmonic oscillator basis.

Alternatively if one considers instead the bound state wave function and
its expansion
\begin{eqnarray}
  \Psi_{B}(r) &=& A_S\,\frac{e^{-\gamma_p}}{r} =
  (A_S^{\rm LO} + \delta A_S)\,\frac{e^{-(\gamma_p^{\rm LO} + \delta \gamma_p) r}}{r}
  \nonumber \\
  &=& \Psi_B^{\rm LO}(r) + \delta \Psi_B(r) \, ,
\end{eqnarray}
with $A_S = \sqrt{2\gamma}\,a_S$ its asymptotic normalization, then
the relative size of the subleading correction is
\begin{eqnarray}
  \frac{\delta \Psi_B}{\Psi_B^{\rm LO}} \to
  \frac{\delta A_S}{A_S^{\rm LO}} + e^{-\delta \gamma_p\,r} \, .
\end{eqnarray}
While the wave function correction originating from the asymptotic
normalization follows the size expectations derived from power
counting, this is not the case for the change in the wave number.
In the infrared ($r \to \infty$) this subleading order effect becomes
arbitrarily large for $\delta\gamma_p < 0$, which violates
power counting expectations.

Here there is a connection with secular perturbation theory, with the only
difference being that the failure happens at large distances instead of
large timescales.
Though if one explicitly considers the time-dependent bound state wave
function, then it will be clear that there is also a secular
mismatch of the time-dependent phase.
The solution is to redefine the secular parameter, in this case the wave
number of the bound state.
In practice this coincides with the well-known
proposal of~\cite{Phillips:1999hh},
which aimed to accelerate the EFT convergence of the bound state properties.

Besides fitting the couplings to the pole~\cite{Phillips:1999hh},
the other two EFT-equivalent solutions discussed here are:
\begin{itemize}
\item[(i)] Choose $r_c = R$ such that the location of the pole is
  reproduced at ${\rm LO}$, while the ${\rm LO}$ coupling is still
  calibrated to reproduce the scattering length.
  For the delta-shell regulator this leads to the equation
  \begin{eqnarray}
    \frac{a_0}{R (R - a_0)} = -\gamma_p\,\left( 1 + {\rm coth}(\gamma_p R)
    \right) \, ,
  \end{eqnarray}
  for which there will be a solution for $R$ if the effective range
  is positive and $a_0$ and $\gamma_p$ have the same sign.
  This is in line with the idea of the cutoff as a convergence dial.
\item[(ii)] Retrofit the ${\rm LO}$ counterterm as to reproduce 
  the ${\rm N^nLO}$ calculation of the pole position~\footnote{That is,
    one begins with a ${\rm LO}$ calculation calibrated to
    the scattering length, calculates the location of the pole at ${\rm NLO}$,
    redefines the ${\rm LO}$ as to reproduce the ${\rm NLO}$ calculation
    of the pole, and so on until reaching the desired order.}.
  This only generates higher order contributions (${\rm N^{n+1}LO}$) to
  the ERE parameters.
  Basically this implements a (sort of) secular perturbation theory
  for the location of the pole by means of a minimal improved
  action that only redefines the ${\rm LO}$ coupling
  a posteriori once the ${\rm N^{n}LO}$ has been computed.
\end{itemize}
The second method might be applied to the case in which there is a transition
from a bound to a virtual state (or vice versa), which has been studied
for instance in~\cite{Yang:2023wci}.
While within a two-body context this type of transition is a curiosity,
they happen in few-body pionless~\cite{Contessi:2017rww,Bansal:2017pwn}
and pionful~\cite{Yang:2020pgi} EFT calculations.

\section{Comparing the EFT and $\hbar$ expansions}
\label{sec:hbar}

There is a really intriguing inconsistency discovered by Epelbaum
et al.~\cite{Epelbaum:2018zli} pertaining the $\hbar$ expansion
of the EFT scattering amplitudes.
As before, auxiliary counterterms offer a solution to this inconsistency and
in the process provide an explanation of it in terms of power counting.

The $\hbar$ expansion is obtained by explicitly considering the powers of
$\hbar$ in the loop expansion of the scattering amplitude:
\begin{eqnarray}
  T = V + \hbar\,V G_0 V + \hbar^2\,V G_0 V G_0 V + \dots \, ,
  \label{eq:T-hbar-expansion}
\end{eqnarray}
where $T$ is calculated in a contact-range EFT with a cutoff $\Lambda$ (and
it has a well-defined $\Lambda \to \infty$ limit).
The inconsistency arises when comparing the power series
around $\hbar \to 0$ and the $\Lambda \to \infty$ limit.
That is, one is confronted with a problem of non-commutativity of
limits and Taylor series.

Before dealing with the inconsistency, it is important to notice that
the previous expansion only makes sense if $\hbar$ is dimensionless,
at least for the specific dimensional choices used in this work
(and in~\cite{Epelbaum:2018zli}) for the scattering
amplitude, the potential and the loop function.
In the physical world $\hbar$ is dimensionful though.
The point is that in the $\hbar$ expansion of Eq.~(\ref{eq:T-hbar-expansion})
the role that $\hbar$ is playing is that of a labeling device
for counting the number of loops.
This observation plays an important role in the explanation of
the inconsistency and its solution.

\subsection{The inconsistency}

For a simple illustration of the inconsistency, one begins by considering
the on-shell scattering amplitude
\begin{eqnarray}
  \frac{1}{k\,\cot{\delta} - i \hbar k} =
  - \frac{M_N}{4\pi} \langle k | T(k) | k \rangle \, ,
\end{eqnarray}
which in a contact-range EFT regularized with PDS takes the form
\begin{eqnarray}
  \frac{1}{k\,\cot{\delta} - i \hbar k} =
  -\frac{\frac{M_N}{4 \pi}\,\sum C_{2n}(\Lambda) k^{2n}}
  {1 + \hbar\,\frac{M_N}{4 \pi} (\Lambda + i k)\,\sum C_{2n}(\Lambda) k^{2n}}
  \, , \nonumber \\
\end{eqnarray}
where $\Lambda$ is the PDS regularization scale (or cutoff).
It is important to stress here that I will treat the PDS regularization
scale $\Lambda$ as a cutoff: the interpretation of the PDS regularization
scale is usually different to that of a cutoff, particularly in what
regards its expected size.
Subtraction constants are expected to be light scales, while cutoffs
are expected to become hard scales.
But, as previously argued, cutoffs behave as light scales
from a formal point of view.
Thus, there is in principle no fundamental difference between
the PDS regularization scale and a cutoff.
If one is not comfortable with this interpretation, one might consider
the previous amplitude as obtained in a fictitious cutoff
regularization scheme in which the loop integrals
coincide with the ones in PDS regularization.

While Epelbaum et al.~\cite{Epelbaum:2018zli} originally uncovered
the inconsistency with cutoff regularization, it also appears
in PDS within the previous interpretation of
its regularization scale.
Yet, PDS allows for a more transparent analysis of the problem,
which is the reason why I adopt it here.
Notice that no power counting considerations enter
in the derivations that follow (though they enter at a later stage,
when reinterpreting this inconsistency from the EFT perspective).

If one now calculates this scattering amplitude with the first two couplings
$C_0$ and $C_2$ and calibrates them to reproduce the scattering length
and effective range, respectively, one obtains:
\begin{eqnarray}
  \frac{1}{k\,\cot{\delta} - i \hbar k} =
  -\frac{1}
  {\frac{1}{a_0}\,\frac{{( 1 - \hbar \Lambda a_0 )}^2}{(1-\hbar \Lambda a_0) + \frac{1}{2} {a_0 r_0 k^2} } + \hbar\, (\Lambda + i k)} \, .
  \nonumber \\ \label{eq:T-matrix-naive}
\end{eqnarray}
This expression has a well-defined infinite cutoff limit given by
\begin{eqnarray}
  \frac{1}{k\,\cot{\delta} - i \hbar k} &=&
  \frac{1}
       {-\frac{1}{a_0} + \frac{1}{2} r_0 k^2 - i \hbar k} \nonumber \\
       &-& \frac{1}{\hbar \Lambda}\,
       {\left( \frac{\frac{1}{2} r_0 k^2}
         {-\frac{1}{a_0} + \frac{1}{2} r_0 k^2 - i \hbar k} \right)}^2
       \nonumber \\
       &+& \mathcal{O}\left( \frac{1}{(\hbar \Lambda)^2}\right) \, .
\end{eqnarray}
In contrast, if one expands in powers of $\hbar$, one arrives to
\begin{eqnarray}
  && \frac{1}{k\,\cot{\delta} - i \hbar k} =
  - a_0 \left( 1 + \frac{1}{2}\,a_0\,r_0\,k^2 \right) \nonumber \\
  && \quad + \hbar \left[ {\left( \frac{1}{2} a_0^2 r_0 k^2 \right)}^2\,\Lambda + i k a_0^2
    {\left( 1 + \frac{1}{2}\,a_0\,r_0\,k^2 \right)}^2 \right] \nonumber \\
 && \quad + \mathcal{O}(\hbar^2) \, ,
\end{eqnarray}
where positive powers of the cutoff $\Lambda$ appear in the expansion.
That is, despite the existence of the infinite cutoff limit, the scattering
amplitude is not renormalizable in an order-by-order basis.
This is the inconsistency pointed out in~\cite{Epelbaum:2018zli}.

\subsection{Solving the inconsistency}

It turns out that this unwanted divergence in the $\hbar$ expansion
can be averted by means of an auxiliary counterterm~\cite{Valderrama:2019yiv},
in particular with:
\begin{eqnarray}
  C_4(\Lambda) = a_0\,\hbar\,\Lambda\,\frac{C_2^2(\Lambda)}{C_0(\Lambda)} \, ,
\end{eqnarray}
which does not carry new physical information beyond what is already contained
in $C_0$ or $C_2$.

However this new counterterm only removes the inconsistency up to $\hbar$,
with it reappearing at $\hbar^2$.
In fact, recalculating the scattering amplitude and reexpanding
in powers of $\hbar$:
\begin{eqnarray}
  && \frac{1}{k\,\cot{\delta} - i \hbar k} =
  - a_0 \left( 1 + \frac{1}{2}\,a_0\,r_0\,k^2 \right) \nonumber \\
  && \quad + 
    i \hbar k \, a_0^2
    {\left( 1 + \frac{1}{2}\,a_0\,r_0\,k^2 \right)}^2  \nonumber \\
    && \quad + \hbar^2
    \left[ {\left( \frac{1}{2} a_0^2 r_0 k^2 \right)}^3\,\Lambda^2 + k^2 a_0^3
    {\left( 1 + \frac{1}{2}\,a_0\,r_0\,k^2 \right)}^3 \right] \nonumber \\
 && \quad + \mathcal{O}(\hbar^3) \, ,
\end{eqnarray}
clearly shows an unmatched divergence at order $\hbar^2$.
This divergence is eliminated by including the auxiliary counterterm
\begin{eqnarray}
  C_6(\Lambda) = {\left( a_0 \hbar \Lambda \right)}^2\,\frac{C_2^3(\Lambda)}{C_0^2(\Lambda)} \, ,
\end{eqnarray}
but the divergence appears again, just at order $\hbar^3$ this time
\begin{eqnarray}
  && \frac{1}{k\,\cot{\delta} - i \hbar k} =
  - a_0 \left( 1 + \frac{1}{2}\,a_0\,r_0\,k^2 \right) \nonumber \\
  && \quad + 
    i \hbar k \, a_0^2
    {\left( 1 + \frac{1}{2}\,a_0\,r_0\,k^2 \right)}^2  \nonumber \\
    && \quad + 
    \hbar^2 k^2 \, a_0^3
    {\left( 1 + \frac{1}{2}\,a_0\,r_0\,k^2 \right)}^3  \nonumber \\
    && \quad + \hbar^3
    \left[ {\left( \frac{1}{2} a_0^2 r_0 k^2 \right)}^4\,\Lambda^3 -i k^3 a_0^4
    {\left( 1 + \frac{1}{2}\,a_0\,r_0\,k^2 \right)}^4 \right] \nonumber \\
 && \quad + \mathcal{O}(\hbar^4) \, .
\end{eqnarray}

Yet, the divergence pattern is now evident, where its solution is the
inclusion of the family of auxiliary counterterms
\begin{eqnarray}
  C_{2n}(\Lambda) = {\left( a_0\,\hbar\,\Lambda \right)}^{(n-1)}\,
  \frac{C_2^n(\Lambda)}{C_0^{n-1}(\Lambda)} \quad \mbox{for $n \geq 2$, }
\end{eqnarray}
which guarantees an $\hbar$ expansion free of divergences
\begin{eqnarray}
    && \frac{1}{k\,\cot{\delta} - i \hbar k} =
  - a_0 \left( 1 + \frac{1}{2}\,a_0\,r_0\,k^2 \right)\, \nonumber \\
  && \quad \times \Big[ 1 - \sum_{n=1}^{\infty} 
    {\left( - i \hbar k \, a_0
      {\left( 1 + \frac{1}{2}\,a_0\,r_0\,k^2 \right)} \right)}^n \Big]
  \nonumber \\
  && \quad = \frac{1}{-\frac{1}{a_0 \left( 1 + \frac{1}{2}\,a_0\,r_0\,k^2 \right)} - i \hbar k} \, . \label{eq:T-matrix-weak}
\end{eqnarray}

This is a really interesting result: it coincides with what one expects
from an EFT in which one expands $\tan{\delta}$ up to $k^2$
\begin{eqnarray}
  \tan{\delta} = - a_0 k \, \left( 1 + \frac{1}{2}\,a_0\,r_0\,k^2 \right) \, .
\end{eqnarray}
This is suspicious: this is a theory in which $\tan{\delta}$ is small
(and thus expansible at low energies), that is, a theory
with {\it weak} scattering.

It is interesting to notice that this is different to the $\Lambda \to \infty$
limit one obtains from iterating the $C_0$ and $C_2$ couplings
in Eq.~(\ref{eq:T-matrix-naive}), that is:
\begin{eqnarray}
  k\,\cot{\delta} = -\frac{1}{a_0} + \frac{1}{2}\,r_0\,k^2 \, ,
\end{eqnarray}
which is what one expects in a theory with {\it strong} scattering.
At first sight this seems to contradict the idea that the effects of
auxiliary counterterms are not observable, as their addition leads to
the T-matrix of Eq.~(\ref{eq:T-matrix-weak}).
But then it is easy to notice that this is just the
$\Lambda \to 0$ limit of Eq.~(\ref{eq:T-matrix-naive}),
which corresponds to weak scattering.

That is, though auxiliary counterterms do not encode physical information,
they do still contain information about the power counting and
will thus force a cutoff independent solution
that is compatible with the counting.

The key observation of why one is actually expanding on $\tan{\delta}$
instead of the naively expected $\cot{\delta}$ is that there were
implicit assumptions about the power counting of $C_0$ and $C_2$
that are not obvious in the first place.
Indeed, the implicit assumption leading to the inconsistency and its solution
is that the $\hbar$ expansion of $C_0$ and $C_2$ begins at $\hbar^0$:
\begin{eqnarray}
  C_{0} = \sum_{n=0}^{\infty} C_0^{[n]} \hbar^n \quad \, \quad
  C_{2} = \sum_{n=0}^{\infty} C_2^{[n]} \hbar^n \, ,
\end{eqnarray}
where owing to the specifics of the regularization used (PDS), it turns
out that $C_0^{[n]} = 0$ and $C_2^{[n]} = 0$ for $n \neq 0$, greatly
simplifying the $\hbar$ expansion of the couplings.

\subsection{Power counting and the $\hbar$ expansion}

For understanding the interplay of the $\hbar$ expansion and power counting,
one might consider the T-matrix in a contact-range theory where only
the $C_0$ coupling is included:
\begin{eqnarray}
  T(k) = \frac{1}{\frac{1}{C_0(\Lambda)} + \hbar\,\frac{M_N}{4 \pi} (\Lambda + i\,k)} \, ,
\end{eqnarray}
which is again derived with PDS regularization.
By calibrating $C_0$ to the scattering length, one obtains the cutoff
independent expression
\begin{eqnarray}
  T(k) = \frac{4\pi}{M_N} \,\frac{1}{\frac{1}{a_0} + i \hbar k} \, ,
\end{eqnarray}
with for $a_0 > 0$ has a bound state at $k = i \gamma$ with
\begin{eqnarray}
  \hbar\,\gamma = \frac{1}{a_0} \, ,
\end{eqnarray}
or, in terms of its two-body binding energy
\begin{eqnarray}
  M_N\,B_2 = - \gamma^2 = - \frac{1}{(\hbar\,a_0)^2} \, ,
\end{eqnarray}
which emphasizes that, if $a_0$ is the quantity to be calibrated
within the EFT, then the two-body system collapses
in the $\hbar \to 0$ limit.
Physically this makes sense because in the classical limit it is not possible
to have closed orbits in regions in which the potential is zero, which
for a genuine contact-range potential comprises all of
the space except the origin.

Yet, within the EFT framework the aforementioned binding energy
can also be an input of the theory (instead of an output).
It is thus always possible to calibrate a given contact-range coupling
to reproduce this input.
Moreover, a bound state is a genuinely quantum effect: its generation
requires the complete resummation of the $\hbar$ expansion.
This is also evident from the equations above: provided the bound state
binding energy or wave number is known, they implicitly assume
that $a_0$ is of order $\hbar^{-1}$: otherwise the equations
will be inconsistent with respect to its $\hbar$ expansion.
That is, the assumption of the existence of a bound state (that can be
described within EFT) contains the implicit assumption that
\begin{eqnarray}
  a_0 = a_0^{[-1]}\,\frac{1}{\hbar} + a_0^{[0]} + a_0^{[1]}\,\hbar + \dots \, .
\end{eqnarray}
which translates into the following $\hbar$ expansion for the $C_0$ coupling:
\begin{eqnarray}
  C_0 = C_0^{[-1]}\,\frac{1}{\hbar} + C_0^{[0]} + C_0^{[1]}\,\hbar + \dots \, ,
\end{eqnarray}
from which one obtains a two-body binding energy
\begin{eqnarray}
  M_N\,B_2 = - \frac{1}{(a_0^{[-1]})^2} +
    \frac{2\,a_0^{[0]}}{(a_0^{[-1]})^3}\,\hbar + \dots \, ,
\end{eqnarray}
that is well-defined in the $\hbar \to 0$ limit.
No judgment is made here regarding the physical interpretation of
the $\hbar \to 0$ limit, only about the formal consistency of
the resulting EFT description, where the $\hbar$ expansion
was initially introduced as a theoretical device for keeping
track of the loop corrections (but not necessarily to obtain
its classical limit).

Conversely, a contact-range theory in which $a_0$ and $r_0$ are to be iterated
to all orders is a theory in which it is possible to have a resonance
(provided that the effective range is negative).
Just like a bound state, a resonance is a genuinely quantum-mechanical
phenomenon that requires infinite resummations.
If the T-matrix that generates the pole of a given resonance is
\begin{eqnarray}
  T(k) = -\frac{4\pi}{M_N}\,
  \frac{1}{-\frac{1}{a_0} + \frac{1}{2}\,r_0\,k^2 + i \hbar k} \, ,
\end{eqnarray}
this requires in turn that
\begin{eqnarray}
  a_0 = a_0^{[-1]}\,\frac{1}{\hbar} + \dots \quad , \quad r_0 = r_0^{[1]} \hbar +
  \dots \, ,
\end{eqnarray}
for the consistent generation of the pole within the EFT
description~\footnote{In the semiclassical approximation
  the s-wave phase shift is naively (i.e. if one ignores
  the Langer correction~\cite{Langer:1937qr})
  given by~\cite{Berry:1972na}
  \begin{eqnarray}
    \delta_{\rm scl}(k) = \frac{1}{\hbar}\,\int_{R_0}^{\infty}\,dr\,\left[
      \sqrt{k^2 + M_N V(r)} - k \right] - \frac{1}{\hbar} k R_0 \, ,
    \nonumber \\
  \end{eqnarray}
  where $R_0$ is the classical turning point of the potential $V(r)$.
  Thus it is a quantity of order $\hbar^{-1}$, which is not surprising
  owing to its quantum mechanical origin. In fact, as pointed out
  in~\cite{Berry:1972na}, the semiclassical phase can be written
  as $\delta_{\rm scl} = \delta_{\rm cl} / \hbar$, where
  $\delta_{\rm cl}$ only involves classical quantities 
  This in turn implies the proposed $\hbar$ scalings $a_0 \sim \hbar^{-1}$
  and $r_0 \sim \hbar$ (though it should be stressed that
  the semiclassical approximation is not expected
  to be accurate at very low momenta).
}.
For the counterterms this expansion reads
\begin{eqnarray}
  C_0 &=& C_0^{[-1]}\,\frac{1}{\hbar} + C_0^{[0]} + C_0^{[1]}\,\hbar + \dots \, , \\
  C_2 &=& C_2^{[-1]}\,\frac{1}{\hbar} + C_2^{[0]} + C_2^{[1]}\,\hbar + \dots \, ,
\end{eqnarray}
and by keeping the $\hbar^{-1}$ terms only, one obtains
the following T-matrix at finite cutoff
\begin{eqnarray}
  T(k) = \frac{4\pi}{M_N}\,
  \frac{1}{\hbar \frac{{(\frac{1}{a_0} - \Lambda)}^2}{(\frac{1}{a_0} - \Lambda) + \frac{1}{2}\,r_0\,k^2} + \hbar\,(\Lambda + i\,k)} \, .
\end{eqnarray}
which is of order $\hbar^{-1}$ and where for simplicity the $\hbar$ superindices
have been removed from $a_0$ and $r_0$ (or alternatively, $a_0^{[-1]}$
and $r_0^{[1]}$ have been renamed $a_0$ and $r_0$).

Owing to the T-matrix now being a purely $\hbar^{-1}$ object there is no
room for inconsistency between the $\hbar$ and $1/\Lambda$ expansions.
Indeed, the second expansion is now trivially enclosed within the first:
\begin{eqnarray}
  T(k) &=& -\frac{4\pi}{M_N}\,
  \frac{1}{\hbar}\Bigg[
    \frac{1}{-\frac{1}{a_0} +  \frac{1}{2}\,r_0\,k^2 - i\,k}
    \nonumber \\
    && \qquad  \quad -\frac{1}{\Lambda}\,
              {\left( \frac{\frac{1}{2}\,r_0\,k^2}{-\frac{1}{a_0} +  \frac{1}{2}\,r_0\,k^2 - i\,k} \right)}^2 \nonumber \\
              && \qquad \quad +
    \mathcal{O}\left( \frac{1}{\Lambda^2} \right) \Bigg] \, , 
\end{eqnarray}
which solves the inconsistency while reproducing exactly
the first two terms in the ERE.

The bottom-line is that within the EFT formalism
the $\hbar$ expansion --- when considered as a loop counting trick,
which is its actual use in Eq.~(\ref{eq:T-hbar-expansion}) ---
is intertwined with power counting:
interactions scaling as $Q^{-1}$ give rise to purely quantum-mechanical
phenomena, such as bound states or resonances, and
hence do implicitly contain a factor of $1/\hbar$.

\subsection{A matter of interpretation?}

Of course, it is natural to ask oneself at this point whether the idea of
the existence of $\hbar^{-1}$ terms in the $\hbar$ expansions of
the couplings is legitimate or not.
After all, the $\hbar$ expansion as applied in EFT
in Ref.~\cite{Epelbaum:2018zli}
is inspired by the quantum effective action, whose $\hbar \to 0$ limit
are the tree diagrams of a given quantum field theory.
Applied to the T-matrix within a contact-range theory this generates
\begin{eqnarray}
  \lim_{\hbar \to 0} T(k) = -\frac{4 \pi}{M_N}\,\sum_{n=0}^{n_{\rm max}} C_{2n} k^{2n} =
  -\frac{4 \pi}{M_N}\,\frac{\tan{\delta}}{k} \Bigg|_{2 n_{\rm max}} \, , \nonumber \\
\end{eqnarray}
where the expansion of the contacts is matched to the expansion of
$\tan{\delta}$ in powers of the momentum squared up to $k^{2 n_{\rm max}}$.
As a consequence of having a low momentum expansion, there is no room
for generating poles at tree level (only perturbative physics).
This type of $\hbar$ expansion as applied to non-relativistic EFTs results
in variations of the trivial fixed point of the renormalization
group~\cite{Birse:1998dk} once the $\hbar \to 0$ limit is taken.
That is, one invariably ends up with EFTs describing weakly interacting
systems where no interaction has to be iterated.

\section{Perturbative and non-perturbative renormalization}
\label{sec:pert}

Changing the perspective once more, it will be interesting to consider
perturbative and non-perturbative renormalization vis-\`{a}-vis each other.
In~\cite{Epelbaum:2018zli} the original inconsistency in the $\hbar$ expansion
is interpreted as a proxy for what is perceived as a more general
inconsistency between perturbative and non-perturbative
renormalization.
There it is pointed out that the mismatch between them is intimately
related to the fact that the amplitudes obtained in the non-perturbative
renormalization process are non-analytic functions of the coupling
constant~\cite{Beane:2000wh,Flambaum:1999zza}.
Hence, the authors of~\cite{Epelbaum:2018zli} write 
that {\it ``the resulting non-perturbative amplitude cannot be expanded
  perturbatively in powers of the coupling constant''}.

This statement is true when applied to the exact non-perturbative amplitude
of singular potentials in the limit in which the cutoff is removed,
which is where actual non-analyticities appear.
This relies on the implicit assumption of the non-commutativity of two limits:
the limit in which the infinite terms of the perturbative expansion are
summed and the $r_c \to 0$ limit.
In this situation inconsistencies similar in nature to the ones already
discussed in~\cite{Epelbaum:2018zli} are bound to appear~\footnote{Conversely,
  one might choose a prescription for the order in which the limits should
  be taken. This is often the case in other areas of physics where
  non-analyticities arise (e.g. phase transitions, for which
  the thermodynamic limit does not converge uniformly).
  Thus, if one takes into account that perturbative and non-perturbative
  renormalization might be interpreted as different infrared fixed points
  of the renormalization group, it is probably acceptable
  for the non-perturbative calculation to not be reducible to
  the perturbative one in the limit of infinite
  separation of scales ($Q/M \to 0$).
}.

The commutativity property of the limits can only be guaranteed if
the perturbative expansion converges {\it uniformly} even for zero
or infinite cutoffs, depending on whether one works
in configuration or momentum space.
But for singular potentials there is a discontinuity between the attractive
and repulsive cases, which is forced by the renormalization process:
in the former counterterms are absolutely required to determine
the low energy solutions, while in the latter the effect
of counterterms ends up being trivial
when the cutoff is removed.

On the contrary, if one substitutes {\it exact} by {\it arbitrarily close}
and if the cutoff is not removed, then the perturbative
expansion of a non-perturbative amplitude will be perfectly
possible for singular potentials.

\subsection{Convergence of the perturbative series}

Here I will show by means of a concrete calculation that
the perturbative series of a singular potential
can indeed be convergent for every finite value of the cutoff
(at least at low external momenta),
even though each individual order in the perturbative expansion
is divergent in the hard cutoff limit.
To illustrate this idea one might consider a two-body system where
the long-range potential is a van der Waals:
\begin{eqnarray}
  V_L(r) = -\frac{C_6}{r^6} = - \frac{1}{M_N}\frac{R_L^4}{r^6} \, ,
\end{eqnarray}
and the short-range potential is modeled
as a boundary condition at the origin.
If one calibrates the short-range boundary condition or counterterm to
reproduce the scattering length, the effective range
is given by~\cite{Flambaum:1999zza}
\begin{eqnarray}
  r_0 = \frac{16\, \Gamma(\frac{5}{4})^2}{3 \pi} R_L -
  \frac{4}{3}\,\frac{R_L^2}{a_0} + \frac{4\, \Gamma(\frac{3}{4})^2}{3 \pi}\,\frac{R_L^3}{a_0^2} \, , \label{eq:r0-vdW}
\end{eqnarray}
which is non-analytic in terms of the coupling constant $C_6$,
from which $R_L$ is derived.

The interesting observation here is that the perturbative series
for the effective range can also be calculated analytically
for a given cutoff radius.
The effective range is given by the integral of
\begin{eqnarray}
  r_0 = 2\,\int_0^{\infty}\,dr\,\left[ {\left( 1-\frac{r}{a_0} \right)}^2 -
    u_0^2(r)\right] \, , \label{eq:effective-range-integral}
\end{eqnarray}
where $a_0$ is the scattering length and $u_0$ the zero-energy wave function.
For simplicity I will take $a_0 \to \infty$ from now on:
the general case with arbitrary scattering length can be derived
following the same steps, but calculations become considerably
more simple with this assumption.
In this case the perturbative series of the
zero energy wave function is given by
\begin{eqnarray}
  u_0(r) = 1 + \sum_{n=1}^{\infty} \hbar^n u_0^{[n]}(r) \, , \label{eq:u0-exp}
\end{eqnarray}
where each term can be calculated iteratively
\begin{eqnarray}
  u_0^{[n+1]}(r) = \int_0^{\infty}\,dr'\,G_0(r,r')\,M_N V_L(r')\,u_0^{[n]}(r') \, ,
  \nonumber \\
\end{eqnarray}
with $G_0(r,r') = (r-r')\,\theta(r'-r)$ the Green function
and $u_0^{[0]}(r) = 1$ for the $a_0 \to \infty$ case.
The perturbative contributions to the wave function can be calculated
recursively, yielding:
\begin{eqnarray}
  u_0^{[n]}(r) = \frac{(-1)^n \Gamma(\frac{5}{4})}{2^{4n}\,\Gamma(1+n)\,\Gamma(\frac{5}{4} + n)}\,{\left(\frac{R_L}{r}\right)}^{4n} \, , \label{eq:un-exp}
\end{eqnarray}
where, owing to the large $n$ behavior of the coefficients within $u_0^{[n]}$,
the perturbative series is convergent for all radii~\footnote{This is
  more specific than what has been previously assumed
  in the literature~\cite{Beane:2000wh}, namely that the
  perturbative series is unable to capture the non-analytic behavior
  of the wave function for radii in which its oscillatory
  nature becomes apparent.}
and its sum is
\begin{eqnarray}
  u_0(r) = \sqrt{\frac{1}{\hbar^{1/4}}\,\frac{2 r}{R_L}}\,\Gamma(\frac{5}{4})\,
  J_{1/4}(\sqrt{\hbar}\,\frac{R_L^2}{2 r^2}) \, , \label{eq:u0-vdW}
\end{eqnarray}
with $J_{\nu}(x)$ denoting the Bessel function of the first
kind~\footnote{Once
  the wave function is known it is easy to calculate the $\hbar$
  expansion of the coupling $C_0$, which for a delta-shell regulator
  is given by
  \begin{eqnarray}
    \frac{M_N\,C_0(r_c)}{4\pi\,r_c^2} = \frac{1}{\hbar}\,\left[
      \frac{u_0'(r_c)}{u_0(r_c)} - \frac{1}{r_c} \right] \, ,
  \end{eqnarray}
  where the inverse factor of $\hbar$ is required if $C_0(r_c)$
  is in units of $[\mbox{energy} \times (\mbox{momentum})^{-3}]$.
  From plugging in the reduced wave function for the van der Waals
  potential in the unitary limit and expanding at long distances
  (or, equivalently, weak coupling), one arrives at
  \begin{eqnarray}
    \frac{M_N\,C_0(r_c)}{4\pi\,r_c^2} = -\frac{1}{\hbar}\,\frac{1}{r_c}
    + \frac{R_L^4}{5\,r_c^5} + \hbar\,\frac{R_L^8}{10\, r_c^9} + \dots \, ,
  \end{eqnarray}
  which begins at $\hbar^{-1}$ and only converges before the first zero
  of the wave function is reached at $(r_c / R_L) \approx 0.424$.
}.

The convergence of the perturbative series for singular potentials is
not a surprise: making use of secular perturbation theory,
Gao~\cite{Gao:1998zza} showed that for an attractive
$1/r^6$ potential the expansion converges 
for $k R_L \leq 1.24\,$~\footnote{Gao originally defines convergence
  in terms of the parameter $\Delta = (k R_L)^2/16$, which requires
  $\Delta \leq 9.65\cdot 10^{-2}$ for s-wave, as listed in
  Table I of~\cite{Gao:1998zza}.}.
Birse~\cite{Birse:2005um} on the other hand extended the results
from Gao~\cite{Gao:1998aaz} for the repulsive $1/r^3$ potential
to the attractive and repulsive tensor one pion exchange
interaction in the chiral limit, leading again to convergence
below a certain value of the momentum.
That is, convergence is expected at low energies.

If one cuts the reduced wave function below $r < r_c$,  the perturbative
series for the effective range can be calculated analytically:
\begin{eqnarray}
  r_0(r_c) = 2\,r_c - 2\,\hbar^{1/4}\,R_L\,\sum_{n=1}^{\infty}
  \frac{c_n}{4 n -1}\,
       {\left(\hbar^{1/4}\,\frac{R_L}{r_c} \right)}^{4n-1} \, ,
       \nonumber \\
       \label{eq:r0-rc}
\end{eqnarray}
where the $c_n$ coefficients are given by
\begin{eqnarray}
  c_n = \frac{9 (-1)^n\,\Gamma(-\frac{3}{4})^2\,\Gamma(\frac{3}{4}+n)}{
  64 \sqrt{2}\, \pi \,\Gamma(\frac{5}{4}+n)\,\Gamma(2+2n)} \, .
\end{eqnarray}
These coefficients behave as $1/((2n)^{(2n+2)})$ for large $n$, which implies
that the power series above converges {\it uniformly}~\footnote{One can
  intuitively understand that the convergence of this series is uniform
  by a direct comparison with the Taylor expansion of $e^{x}$, for which
  the $n$-th order coefficient is $1/n!$ and thus behaves as $1/n^n$.
  This is to be compared with $1/((2n)^{(2n+2)})$ for the series of
  the effective range with van der Waals forces.
  It is a well-known fact that for $|x| < x_0$, with $x_0 > 0$ arbitrary
  but finite, the Taylor series of the exponential function
  converges uniformly. From this, it is apparent that a series with
  coefficients that decrease in size even faster than
  for the exponential should also converge uniformly.
}
for every finite value of $x = (\hbar^{1/4} R_L/r_c)$.
Thus this power series converges to a continuous function of $x$, which
can indeed be calculated
\begin{eqnarray}
  \sum_{n=1}^{\infty} \frac{c_n}{4 n - 1} x^{4n-1} = \frac{1}{x}\,\left[
  1 - {}_{1}F_{2}(-\frac{1}{4}; \frac{5}{4}, \frac{3}{2}; - \frac{x^4}{4})
  \right] \, , \nonumber \\
  \label{eq:sum-hypergeometric}
\end{eqnarray}
where ${}_p F_q (a_1, \dots , a_p; b_1 , \dots, b_q: z)$ denotes
the generalized hypergeometric function~\footnote{A
  noteworthy aspect of the perturbative treatment of attractive singular
  potentials is that it is surprisingly well-behaved, even when compared
  with quantum-mechanical perturbative expansions that do not require
  renormalization in the first place. For instance, the perturbative
  expansion of the anharmonic oscillator is an asymptotic series
  with a zero convergence radius in terms of the coupling
  constant~\cite{Bender:1969si,Bender:1971gu,Bender:1973rz}.
}.

Yet, a really interesting aspect of the power series of Eq.~(\ref{eq:r0-rc})
for the effective range is that it can approximate the full non-perturbative
and non-analytic result of Eq.~(\ref{eq:r0-vdW}) with arbitrary
precision for finite $r_c > 0$.
This can be deduced from (i) the uniform convergence of the power series
for finite $r_c$ and (ii) the existence of a closed expression for
the sum of this power series (i.e. Eq.~(\ref{eq:sum-hypergeometric}))
that converges to the full result of Eq.~(\ref{eq:r0-vdW})
for $r_c \to 0$.
It is in this sense of {\it arbitrarily close approximation} that
one can indeed expand perturbatively a non-perturbative result.

For $x \to \infty$ ($r_c \to 0$) the series converges to:
\begin{eqnarray}
  \lim_{x \to \infty} \sum_{n=1}^{\infty} \frac{c_n}{4 n -1}\,x^{4n-1}
      = - \frac{8 \Gamma(\frac{5}{4})^2}{3 \pi} \, ,
\end{eqnarray}
as deduced either from a direct comparison with the expected effective range
for $a_0 \to \infty$, Eq.~(\ref{eq:r0-vdW}), or from directly
taking the limit in Eq.~(\ref{eq:sum-hypergeometric}).
Here convergence is point-wise though, which can be deduced
from the fact that the infinite sum and
the limit do not commute.
This isolates the origin of non-analytic behavior in terms of
the coupling constant to the $r_c \to 0$ limit.
The conclusion is that the non-perturbative effective range
cannot be reproduced {\it exactly} by means of a
perturbative expansion.
In this second sense Ref.~\cite{Epelbaum:2018zli} is right.

Of course the cutoff dependence of the effective range can be stabilized
by including auxiliary counterterms.
The explicit contribution to the effective range from a perturbative,
energy-dependent counterterm (with a delta-shell regulator) is 
\begin{eqnarray}
  \frac{\Delta\,r_0}{2} &=& M_N\,\int_0^{\infty}\,dr\,
  \,\frac{C_2(r_c)}{4\pi r_c^2}\,\delta(r-r_c)\,
  u_0^2(r)
  \nonumber  \\
  &=& 
  M_N\,\frac{C_2(r_c)}{4\pi r_c^2}\,u_0^2(r_c) \, ,
\end{eqnarray}
where the expression in the first line is derived from the application
of distorted wave perturbation theory
(one might consult~\cite{PavonValderrama:2025azr}
for a recent EFT-centric, r-space exposition).
By choosing $\Delta r_0 = r_0 - r_0(r_c)$ with $r_0$ and $r_0(r_c)$ as given
by Eqs.~(\ref{eq:r0-vdW}), ~(\ref{eq:r0-rc}) and (\ref{eq:sum-hypergeometric}),
one guarantees the exact reproduction of the van der Waals effective range.
If one then expands $u_0(r_c)$ and $r_0(r_c)$ perturbatively,
Eqs.~(\ref{eq:u0-exp}), ~(\ref{eq:un-exp}) and~(\ref{eq:r0-rc}),
one obtains the perturbative expansion of $C_2(r_c)$
by matching powers of $R_L/r_c$.
This trivially ensures uniform convergence of the effective range,
with the $r_c \to 0$ point-wise convergence issue
now buried in the power series of $C_2(r_c)$,
which is not observable anyways.

This is basically a reshuffling of the EFT expansion (and perfectly
allowed by the EFT principles): the individual terms in the perturbative
expansion are not observable, only their sum is.
Yet, this reordering ensures that the perturbative expansion of
the effective range is finite at each order (in particular,
every contribution is identically zero beyond the lowest
order in this case).
Thus auxiliary counterterms solve the theoretical tension between the
non-perturbative amplitudes and their perturbative reexpansion,
though their practical implementation depends on prior
knowledge of the full non-perturbative result.

\subsection{Understanding non-perturbative renormalization}

The comparison between the non-perturbative effective range and its
non-perturbative series for van der Waals reveals the following:
\begin{itemize}
\item[(i)] The perturbative series is uniformly convergent for every $r_c > 0$
  (and $0 < R_L < \infty$), but the individual terms in the perturbative
  expansion are divergent for $r_c \to 0$.
\item[(ii)] The renormalization of the perturbative series order-by-order,
  i.e. its perturbative renormalization, misses this global property
  of convergence and ends up having a different counterterm structure.
\end{itemize}
In this example, perturbative renormalizability involves a loss of
information about the perturbative series that does not happen
in the non-perturbative case: in the perturbative renormalization
of the van der Waals potential the effective range is a free
parameter instead of a prediction.
In a sense it could even be argued that non-perturbative renormalization
is more complete or fundamental than the perturbative one, at least
for attractive singular finite-range potentials.

At this point it is interesting to revisit the previous
calculation of a contact-range theory including range corrections,
Eq.~(\ref{eq:T-matrix-naive}), which yields the following expansion
for the tangent of the phase shift
\begin{eqnarray}
  \frac{\tan{\delta}}{k} &=& -a_0\,\left( 1 + \frac{1}{2}\,a_0 r_0 k^2 \right)
   \nonumber \\
   &+& \frac{1}{4}\,a_0^4 r_0^2 k^4 \,\sum_{n=0}^{\infty} {\left( -\frac{a_0}{2}\,
     (a_0 r_0 k^2 - 2)
   \right)}^n
   (\hbar \Lambda)^{n+1} \, , \nonumber \\
\end{eqnarray}
where this series is only convergent if
\begin{eqnarray}
  \left|\frac{a_0}{2}\,( a_0r_0\,k^2 - 2 )\,\hbar\,\Lambda
  \right |\leq 1 \, .
\end{eqnarray}
If this condition is met, the sum of the perturbative series exists
in a rigorous sense and is given by
\begin{eqnarray}
  \frac{\tan{\delta}}{k} &=& -
  \frac{1}{\frac{{(\frac{1}{a_0} - \hbar \Lambda)}^2}{(\frac{1}{a_0} - \hbar\,\Lambda) + \frac{1}{2}\,r_0\,k^2} + \hbar\,\Lambda} \, .
\end{eqnarray}
On the contrary, if the convergence criterion is not met, which is what
happens if one takes the $\Lambda \to \infty$ limit, the previous
expression is not the sum of the perturbative series
but its analytic continuation.
Here it can be argued that though the previous analytic continuation
makes perfect sense at the mathematical level, it might not be
physically valid.

This observation provides a reason for the warning against
the use of hard cutoffs in non-perturbative renormalization
by Epelbaum et al.~\cite{Epelbaum:2018zli},
where the inconsistency in the $\hbar$ expansion is interpreted
as a strong hint that certain types of mathematical
extrapolations are not legitimate.
After all, this expansion is not convergent for hard cutoffs.
From this viewpoint it makes sense to give precedence to the perturbative
renormalization of amplitudes over its non-perturbative counterpart.

Yet, the counterexample of the effective range in the van der Waals potential
shows the existence of cases in which non-perturbative renormalization
might take precedence over its perturbative formulation.
While the individual terms in the perturbative series do not make sense
in the $r_c \to 0$ limit, their non-perturbative sum does.
Specifically, the non-analyticity of observables in the coupling constant
appears naturally (in the sense of an arbitrarily improvable approximation)
as a result of summing the convergent perturbative series.

A pair of final clarifications are in order.
On the one hand, the $1/r^6$ potential is a {\it physically relevant}
finite-range potential that corresponds to the longest range part of
the interaction between neutral atoms.
It does not have to be valid arbitrarily close to the origin
in order to be relevant for the physics of neutral atoms.
On the other, whether non-fractional powers of the coupling constant
do rigorously exist or manifest in nature is ultimately a moot point.
The reason can be traced back to the fact
that for sufficiently small, but finite cutoffs,
one can approximate the non-perturbative amplitudes containing
the non-analyticities in the coupling constant arbitrarily.
In the case at hand --- the effective range of the van der Waals potential ---
the difference between its finite and zero cutoff limit is bounded by
\begin{eqnarray}
  r_0(r_c) - r_0(r_c \to 0)
  &=& 2 \int_0^{r_c}\,dr\,u_0^2(r) \nonumber \\
  &\leq& 2 \, r_c \, ,
\end{eqnarray}
which is obtained by cutting the reduced wave function for $r < r_c$
in Eq.~(\ref{eq:effective-range-integral}) and then considering
that $u_0(r)\leq 1$ for the van der Waals potential and
the infinite scattering length limit.
For $r_c \lesssim R_L$ there is a tighter bound 
\begin{eqnarray}
  r_0(r_c) - r_0(r_c \to 0)
  &\leq& \mathcal{O}(r_c^4) \, ,
\end{eqnarray}
which is based on the $r/R_L \lesssim 1$ behavior of the non-perturbative
reduced wave function of Eq.~(\ref{eq:u0-vdW}).
This rate of convergence is compatible with the one of $\mathcal{O}(r_c^{n/2+1})$
for power-law singular potentials of the $1/r^n$ type found
in~\cite{PavonValderrama:2007nu}.
In this particular example it is perfectly possible to get arbitrarily
close to the non-analytic expression of Eq.~(\ref{eq:r0-vdW})
with a finite cutoff for which the perturbative series
converges.

Thus, if one accepts the idea that finite cutoffs do model finite
separations of scales effectively, the conclusion is that
the non-analytic expressions obtained with singular
potentials become excellent approximations in
systems with a clean separation of scales.
In the case of neutral atoms, calculations with phenomenological potentials
usually reproduce Eq.~(\ref{eq:r0-vdW}) with great
accuracy~\cite{Cordon:2009wh}.

This discussion is applicable {\it mutatis mutandis} to the status of
singular potentials in nuclear physics.
In~\cite{Epelbaum:2018zli} it is argued that the $1/r^3$ behavior of
the one pion exchange potential has nothing to do with the real world,
where the deeply bound states typical of attractive singular
interactions are nowhere to be seen
in the two-nucleon system.
Besides, this $1/r^3$ behavior is only apparent at distances below
the Compton wavelength of the pion, which apparently gives it
a short-range character.
The counterargument is that the appearance of deeply bound states and their
number are contingent on the breakdown scale of the theory:
while in the two-nucleon system there is no deeply bound state
with the $1/r^3$ potential,
in systems of two alkali atoms it is easy to find several dozens
of deeply bound states with the $1/r^6$ potential.
This issue is immaterial though for their non-perturbative renormalization.
Moreover, in the chiral limit the one pion exchange potential is a pure $1/r^3$
potential at arbitrarily long distances, indicating that this is
a genuine long range property.

Ultimately this discussion is independent of the issues of the consistency of
the non-perturbative renormalization of singular potentials: once there is
consistency (and hence cutoff independence), nothing prevents the use of
a finite cutoff of the order of the breakdown scale if one is not
comfortable with the ontological status of singular
potentials at shorter distances.
Indeed, this might be the best choice of a cutoff: the absence of multiple
bound states in the two-nucleon system signals a poor separation of scales.
This in turn suggests that softer cutoffs might be preferable as they are
more likely to generate observables closer to the physical
ones~\footnote{Yet,
  for a counterexample, the calculation of the deuteron observables
  with the one pion exchange potential once the cutoff
  is removed~\cite{PavonValderrama:2005gu} tends to be
  more accurate than analogous calculations
  with finite cutoffs~\cite{Gezerlis:2014zia,Epelbaum:2014sza}.
  This observation does not extend though beyond the two-nucleon system:
  the calculation of the triton binding energy at ${\rm LO}$ is indeed
  more accurate with finite cutoffs than
  with large cutoffs~\cite{Nogga:2005hy}.} (provided
that the corrections coming from short-range physics shift said
observables in the same direction as finite-cutoff effects,
which seems to be the case, at least in pionless EFT).

The elephant in the room is whether the $r_c \to 0$ limit is well-defined and
what its relation is with the type of inconsistencies explored
in~\cite{Epelbaum:2018zli}.
The approach I have adopted here is rather practical: I simply point out
that for arbitrarily hard (but still finite) cutoffs non-perturbative
renormalization is consistent.
However, I have explicitly avoided taking the limit in which the cutoff
is removed, as this opens further technical problems beyond
the scope of the present manuscript.
Nonetheless, this is important in the sense that if this limit is
well-defined and non-pathological then it is possible to reinterpret
the results of quantum defect
theory~\cite{Gao:1998zza,Gao:1998aaz,Flambaum:1999zza} as
rigorous results in non-relativistic EFT (where the relation
between them has scarcely been explored~\cite{Elhatisari:2013swa}).

\section{Discussion and Conclusions}

I have considered the impact of non-observable contact-range interactions
--- auxiliary or redundant counterterms --- on the non-observable
properties of EFTs.
This might look like a pointless exercise at first sight, yet the internal
consistency of EFTs belongs to the aforementioned
non-observable properties. 
In particular, auxiliary counterterms play a role in:
\begin{itemize}
\item[(i)] Improving the cutoff dependence (and consequently
  the convergence properties) of the EFT description,
  including the complete elimination of the residual
  cutoff dependence (if one chooses to, modulo
  technical limitations that might prevent
  this goal from being achieved).
\item[(ii)] Explaining why apparent inconsistencies within the EFT expansion,
  in particular when they involve comparisons with the hard cutoff
  expansion, are not genuine inconsistencies and instead have 
  simple explanations in terms of neglected
  ingredients of the EFT description.
\end{itemize}

The first of the previous two ideas helps to frame the role of cutoff
dependence and independence within EFT.
They provide a different interpretation to the idea of improved
actions (originally developed in the context of lattice QCD
calculations~\cite{Symanzik:1983dc}) as applied to
EFTs~\cite{Contessi:2023yoz,Contessi:2024vae,Contessi:2025xue}:
after all, including part of the effective range at ${\rm LO}$ is
equivalent to making ${\rm LO}$ calculations with a finite cutoff.
If the effective range can be reproduced in this way, which requires
it to be positive, then the equivalence holds.
Owing to the idea that there is no privileged solution when the cutoff
is removed, the improved actions in~\cite{Contessi:2023yoz,Contessi:2024vae,Contessi:2025xue}
can be reinterpreted as simply making use of the inherent ${\rm LO}$
uncertainty, where the promotion of a fictitious effective range
operator is a device to exploit this freedom.
A closely related use case is the infrared failure of the EFT expansion
around the poles of the scattering amplitude, which has been previously
explored in the literature~\cite{Phillips:1999hh,Contessi:2017rww,Bansal:2017pwn,Yang:2023wci}.
Here, the approaches of~\cite{Phillips:1999hh,Yang:2023wci}
are reinterpreted in the same terms as (or as a special instance of)
the improved actions.

The second use case of the auxiliary counterterms is the understanding and
reinterpretation of one of the intriguing inconsistencies
discovered in~\cite{Epelbaum:2018zli}.
There the observation is made that when one considers the $\hbar$
expansion of the scattering amplitudes calculated within EFT,
then one observes that renormalization is not happening
in an order-by-order basis within the previous
expansion in the hard cutoff limit~\cite{Epelbaum:2018zli},
which is rather puzzling.
This does not pertain the observable properties of EFT --- the individual terms
in the $\hbar$ or cutoff expansions of EFT amplitudes are just part of the
internal machinery of the EFT framework, but not of its predictions ---
and it turns out that the inconsistency can in fact be remedied by
adding a series of operators with no impact
on observable quantities

One lesson from the previous exercise is the realization that
the $\hbar$ and power counting expansions are intertwined: terms that
are of order $Q^{-1}$ in the EFT expansion --- the terms that give rise
to genuine quantum effects such as bound states or resonances ---
are better counted as $\hbar^{-1}$ if one wants the $\hbar$
expansion to reproduce them.

Though in a first instance unrelated to the auxiliary counterterms,
a second potential inconsistency conjectured in~\cite{Epelbaum:2018zli}
is that the appearance of non-analytic
dependence on the coupling constant (commonplace in the non-perturbative
treatment of singular potentials when the cutoff is removed) is probably
incompatible with the $\hbar$ expansion.
Here I provide a counterexample --- the $\hbar$ expansion of the effective
range of the attractive $1/r^6$ potential --- which shows that
\begin{itemize}
\item[(i)] The $\hbar$ (or loop) expansion is convergent for every cutoff
  (and uniformly
  convergent for finite cutoffs) and its sum generates fractional powers of
  the coupling constant, i.e. non-analytic behavior,
  in the $r_c \to 0$ limit (which may be approximated
  with arbitrary precission with finite cutoffs).
\item[(ii)] The renormalized non-perturbative effective range, when expanded
  in $\hbar$, does not generate a renormalized perturbative series (but an
  unrenormalized one instead, in which each term eventually diverges when
  the cutoff is removed).
\end{itemize}
The second observation updates the statement originally made
in~\cite{Epelbaum:2018zli}, that
{\it ``in self-consistent EFTs properly renormalized non-perturbative
expressions, when expanded in $\hbar$, must reproduce
the renormalized perturbative series''} by indicating
that this is not always the case.
The statement is probably derived from the experience in the renormalization of
momentum-independent contact-range potentials and other analytically-solvable
models~\cite{Epelbaum:2009sd}, where it neatly applies.

Yet, a different framing is that this statement acts as a choice
resulting in two distinct conceptions of non-perturbative renormalization.
If one accepts it, the prescription of using moderate cutoffs follows
from the requirement of having a renormalized perturbative series (in the
sense of the finiteness and naturalness of each individual term)
when cutoff independence is not achievable order-by-order.
The distinction might be artificial though: the inclusion of auxiliary
counterterms can make the terms in the perturbative series finite,
implying the equivalence of both formulations.

The bottom-line is that the non-analytic dependence on the coupling constant
does not pose a conceptual challenge to the non-perturbative
renormalization of the one pion exchange
potential~\cite{PavonValderrama:2005gu}, whose tensor component
is singular, and which is the basis for the EFT expansion
proposed in~\cite{Nogga:2005hy} (which advocated the perturbative
treatment of the subleading corrections).
This does not preclude though the existence of other types of inconsistencies
that might eventually be discovered in the EFT expansion, especially when
applied to nuclear physics.
In this regard, the iteration of the one pion exchange potential implies
the identification of $\Lambda_{\rm NN}$ as a low energy scale,
and as explained by Birse~\cite{Birse:2005um}, {\it 
``since this scale is built out of quantities that are treated as
high-energy scales in ChPT, this analysis leaves open the question
of how to make this theory consistent with chiral expansions of
other effective operators''}, where ChPT refers to
chiral perturbation theory.
Two decades after this observation the theoretical grounds for the expansions
of other effective operators appear to be well-understood even when
one pion exchange is iterated~\cite{PavonValderrama:2014zeq},
not to mention that there exist already a few practical
calculations too~\cite{Shi:2022blm,Liu:2022cfd}.

Yet, renormalization in non-perturbative settings remains less understood
than in perturbative ones, which explains why new problems
and solutions are still appearing.
Adopting the cautious approach of~\cite{Epelbaum:2009sd} makes sense,
particularly in the initial exploratory phases.
But caution should be exercised in moderation,
as it should not prevent the adoption of new advances in nuclear EFT
once it becomes apparent that they rest on solid grounds.
The combination of non-perturbative renormalization and singular finite-range
interactions is however amenable to unexpected results
in need of interpretation, of which the ones dealt with
here are but a small sample of what might be eventually
found in the future (with a recent example being
the discovery of the {\it exceptional cutoffs}~\cite{Gasparyan:2022isg},
which has generated a series of analyses and their corresponding
solutions~\cite{Peng:2024aiz,Yang:2024yqv,PavonValderrama:2025zzk,Peng:2025ykg}).

To summarize, because of their non-observable nature auxiliary or redundant
counterterms are a (usually) neglected ingredient of EFT.
They are a powerful analysis tool by which to clarify, diagnose and
then cure internal EFT inconsistencies that have been previously
pointed out in the literature.
Yet, they have a more practical use case in the construction of
improved actions by which to enhance the convergence of EFT
calculations, be it by themselves or by the inclusion
of a fictitious effective range (a theoretical device
which has recently begun to be
investigated~\cite{Contessi:2023yoz,Contessi:2024vae,Contessi:2025xue}).

\begin{acknowledgments}
  I would like to thank Ubirajara van Kolck for lively discussions about
  the interpretations of the cutoff in EFT, Evgeny Epelbaum,
  Ashot Gasparyan and Jambul Gegelia for further discussions and
  their detailed and abundant comments on this manuscript
  that have been incredibly useful, and Mario S\'{a}nchez
  S\'{a}nchez for discussions and noticing a few mistakes.
  I also thank Johannes Kirscher for pointing out that the term
  ``redundant counterterm'' might be amenable to confusion.
  This work is partly supported by the National Natural Science Foundation
  of China under Grant No. 12435007.
\end{acknowledgments}



%

\end{document}